\documentclass{elsart}

\usepackage[square,comma]{natbib}
\usepackage{graphicx}
\usepackage{pxfonts}
\usepackage{lineno}

\usepackage{amssymb}

\journal{}

\begin{document}

\thispagestyle{empty}
\begin{Large}
\textbf{DEUTSCHES ELEKTRONEN-SYNCHROTRON}

\textbf{\large{Ein Forschungszentrum der
Helmholtz-Gemeinschaft}\\}
\end{Large}

DESY 10-053

April 2010

\begin{eqnarray}
\nonumber &&\cr \nonumber && \cr \nonumber &&\cr
\end{eqnarray}
\begin{eqnarray}
\nonumber
\end{eqnarray}
\begin{center}
\begin{Large}
\textbf{A simple method for controlling the line width of SASE
X-ray FELs}
\end{Large}
\begin{eqnarray}
\nonumber &&\cr \nonumber && \cr
\end{eqnarray}

\begin{large}
Gianluca Geloni,
\end{large}
\textsl{\\European XFEL GmbH, Hamburg}
\begin{large}

Vitali Kocharyan and Evgeni Saldin
\end{large}
\textsl{\\Deutsches Elektronen-Synchrotron DESY, Hamburg}
\begin{eqnarray}
\nonumber
\end{eqnarray}
\begin{eqnarray}
\nonumber
\end{eqnarray}
ISSN 0418-9833
\begin{eqnarray}
\nonumber
\end{eqnarray}
\begin{large}
\textbf{NOTKESTRASSE 85 - 22607 HAMBURG}
\end{large}
\end{center}
\clearpage
\newpage

\begin{frontmatter}



\title{A simple method for controlling the line width of SASE X-ray FELs}


\author[XFEL]{Gianluca Geloni\thanksref{corr},}
\thanks[corr]{Corresponding Author. E-mail address: gianluca.geloni@xfel.eu}
\author[DESY]{Vitali Kocharyan}
\author[DESY]{and Evgeni Saldin}

\address[XFEL]{European XFEL GmbH, Hamburg, Germany}
\address[DESY]{Deutsches Elektronen-Synchrotron (DESY), Hamburg,
Germany}

\begin{abstract}

This paper describes a novel single-bunch self-seeding scheme for
generating highly monochromatic X-rays from a baseline XFEL
undulator. A self-seeded XFEL consists of two undulators with an
X-ray monochromator located between them. Previous self-seeding
schemes made use of a four-crystal fixed-exit monochromator in
Bragg geometry. In such monochromator the X-ray pulse acquires a
cm-long path delay, which must be compensated.  For a single-bunch
self-seeding scheme this requires a long electron beam bypass,
implying modifications of the baseline undulator configuration. To
avoid this problem, a double bunch self-seeding scheme based on a
special photoinjector setup was recently proposed. At variance,
here we propose a new time-domain method of monochromatization
exploiting a single crystal in the transmission direction, thus
avoiding the problem of extra-path delay for the X-ray pulse. The
method can be realized using a temporal windowing technique,
requiring a magnetic delay for the electron bunch only. When the
incident X-ray beam satisfies the Bragg diffraction condition,
multiple scattering takes place and the transmittance spectrum in
the crystal exhibits an absorption resonance with a narrow
linewidth. Then, the temporal waveform of the transmitted
radiation pulse is characterized by a long monochromatic wake. The
radiation power within this wake is much larger than the shot
noise power. At the entrance of the second undulator, the
monochromatic wake of the radiation pulse is combined with the
delayed electron bunch, and amplified up to saturation level. The
proposed setup is extremely simple and composed of as few as two
simple elements. These are the crystal and the short magnetic
chicane, which accomplishes three tasks by itself.  It creates an
offset for crystal installation, it removes the electron
micro-bunching produced in the first undulator, and it acts as a
delay line for temporal windowing. Using a single crystal
installed within a short magnetic chicane in the baseline
undulator, it is possible to decrease the bandwidth of the
radiation well beyond the XFEL design down to $10^{-5}$.  The
installation of the magnetic chicane does not perturb the
undulator focusing system and does not interfere with the baseline
mode of operation. We present feasibility study and
exemplifications for the SASE2 line of the European XFEL.
\end{abstract}

%
%
%
\end{frontmatter}



\section{\label{sec:intro} Introduction}

As a consequence of the start-up from shot noise, the longitudinal
coherence of X-ray SASE FELs is rather poor compared to
conventional optical lasers. The coherence time is defined by the
inverse spectral width. For conventional XFELs
\cite{tdr-2006}-\cite{SPRIN} this is typically two orders of
magnitude shorter than the electron pulse duration.  Hence, the
typical XFEL pulse bandwidth is about two orders of magnitude
larger than the Fourier transform limited value for the total
bunch length. Given this physical properties, it is in principle
possible to improve the longitudinal coherence and produce X-rays
with a bandwidth of about $10^{-5}$.

\begin{figure}[tb]
\includegraphics[width=1.0\textwidth]{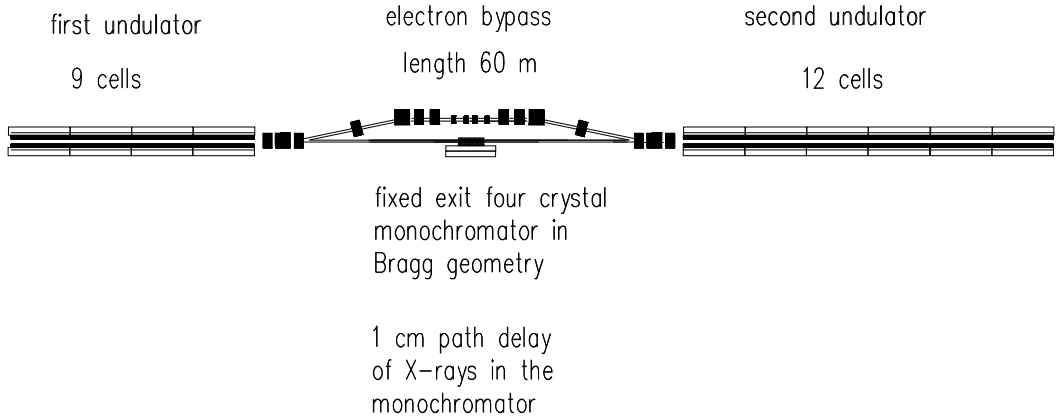}
\caption{Design of an undulator system for the narrow bandwidth
mode of operation. The scheme is based on the use of a single
bunch self-seeding scheme with a fixed-exit four-crystal
monochromator in Bragg geometry. The presence of the monochromator
introduces a path delay of the X-rays, which has to be compensated
with the introduction of a long electron beam bypass.} \label{fd5}
\end{figure}

Self-seeding schemes have been studied to reduce the bandwidth of
SASE X-ray FELs \cite{SELF}-\cite{SCOM}. A self-seeded FEL
consists of two undulators with an X-ray  monochromator located
between them. The first undulator operates in the linear high-gain
regime starting from the shot-noise in the electron beam. After
the first undulator, the output radiation passes through the X-ray
monochromator, which reduces the bandwidth to the desired value,
smaller than the FEL bandwidth. While the radiation is sent
through the monochromator, the electron beam passes through a
bypass, which removes the electron micro-bunching introduced in
the first undulator and compensates for the path delay created
during the passage in the monochromator. At the entrance of the
second undulator, the monochromatic X-ray beam is then combined
with an electron beam and amplified up to the saturation level.
The radiation power at the entrance of the second undulator is
dominant over the shot noise power, so that the bandwidth of the
input signal is smaller than the bandwidth of the FEL amplifier.
The realization of this self-seeding scheme for the European XFEL
requires two undulators, the first $54$ m long ($9$ cells) and the
second $72$ m long ($12$ cells), separated by a four-crystal,
fixed-exit monochromator in reflection (Bragg) geometry. In the
monochromator, the X-ray pulse acquires a centimeter-long path
delay, which must be compensated. For a single bunch self-seeding
scheme this requires a long electron beam bypass with a length of
about $60$ m, implying modifications of the baseline undulator
configuration, Fig. \ref{fd5}. As an attempt to go around this
obstacle, a double-bunch self-seeding scheme was proposed in
\cite{OURL}, based on a photoinjector setup using a laser pulse
doubler \cite{DOUB}.

All X-ray crystal monochromators operate in the frequency domain
as bandpass filters.  In this paper we propose, instead, a new
method of monochromatization based on the use of a single crystal
in the transmission direction as a bandstop filter. In this way,
the problem with the extra path-delay of the X-ray pulse does not
exist at all. Our scheme can be realized with the help of a
temporal windowing technique.  When the incident X-ray beam
satisfies the Bragg diffraction condition, multiple scattering
takes place in the crystal and the spectrum of the reflectance
exhibits a narrow line width. As a consequence, the spectrum of
the transmittance exhibits an absorption resonance with a narrow
line width, thus behaving as a a bandstop filter. After such
bandstop filter, the frequency spectrum of the transmitted X-ray
pulse experiences a strong temporal separation. In particular, the
temporal waveform of transmitted radiation pulse exhibits a long
monochromatic wake.  At the entrance of the second undulator, the
monochromatic wake of the radiation pulse is combined with the
delayed electron bunch, thus exploiting the before-mentioned
temporal windowing technique, and it is amplified up to
saturation. The wake power is dominant over the shot noise power,
so that the bandwidth of the input signal is small, compared to
the bandwidth of the FEL amplifier.

\begin{figure}[tb]
\includegraphics[width=1.0\textwidth]{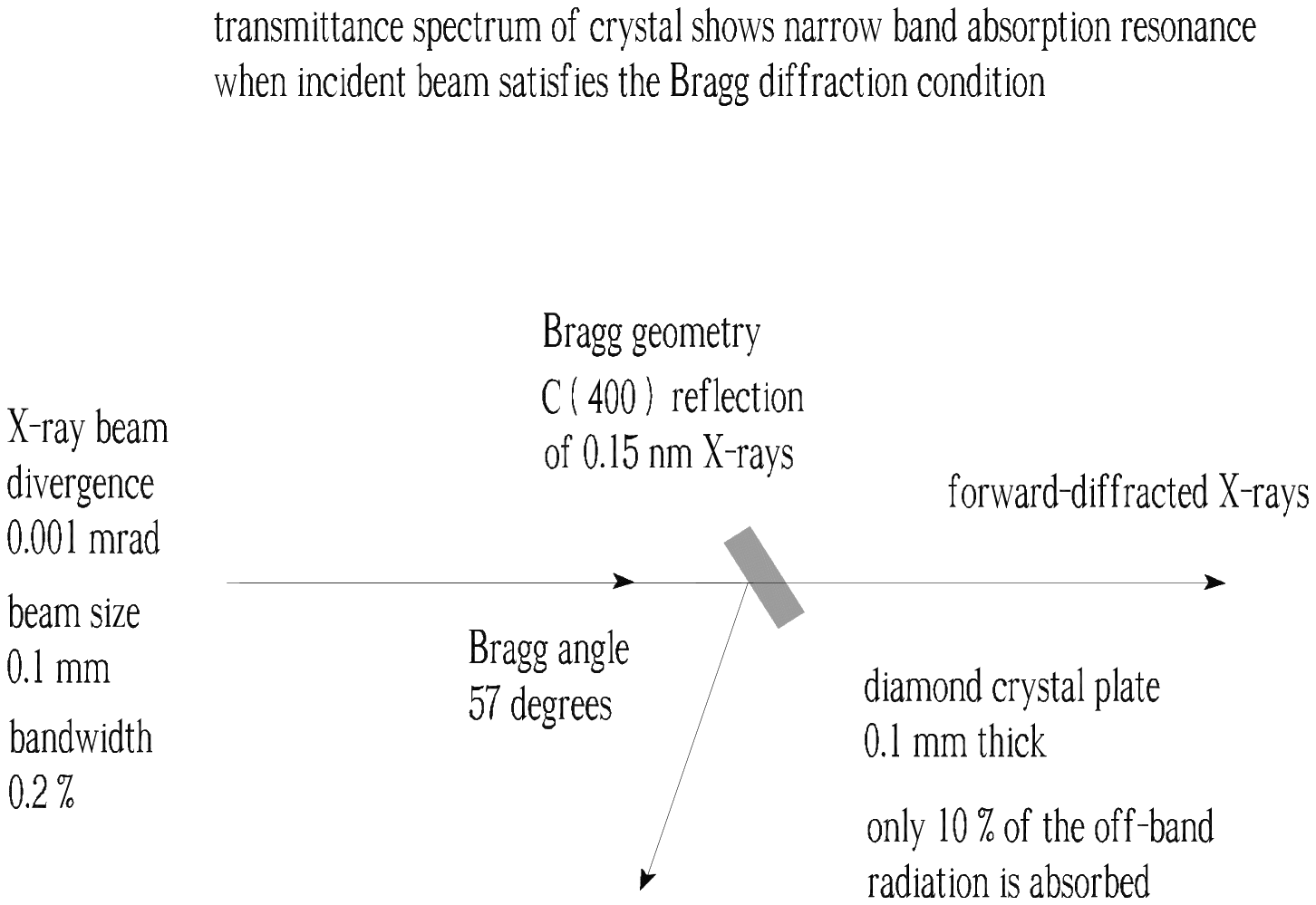}
\caption{Forward diffraction in a single crystal in Bragg
geometry. Due to multiple scattering, the transmittance spectrum
in a crystal shows an absorption resonance with a narrow
($10^{-5}$) linewidth. Resonant wavelength and incident angle of
the X-ray beam satisfy the Bragg diffraction condition. When the
incident angle and the spectral contents of the incoming beam
satisfies the Bragg diffraction condition, the temporal waveform
of the transmitted radiation pulse exhibits a long monochromatic
wake. The duration of the wake is inversely proportional to the
bandwidth of the absorption resonance.} \label{fd4}
\end{figure}
\begin{figure}[tb]
\includegraphics[width=1.0\textwidth]{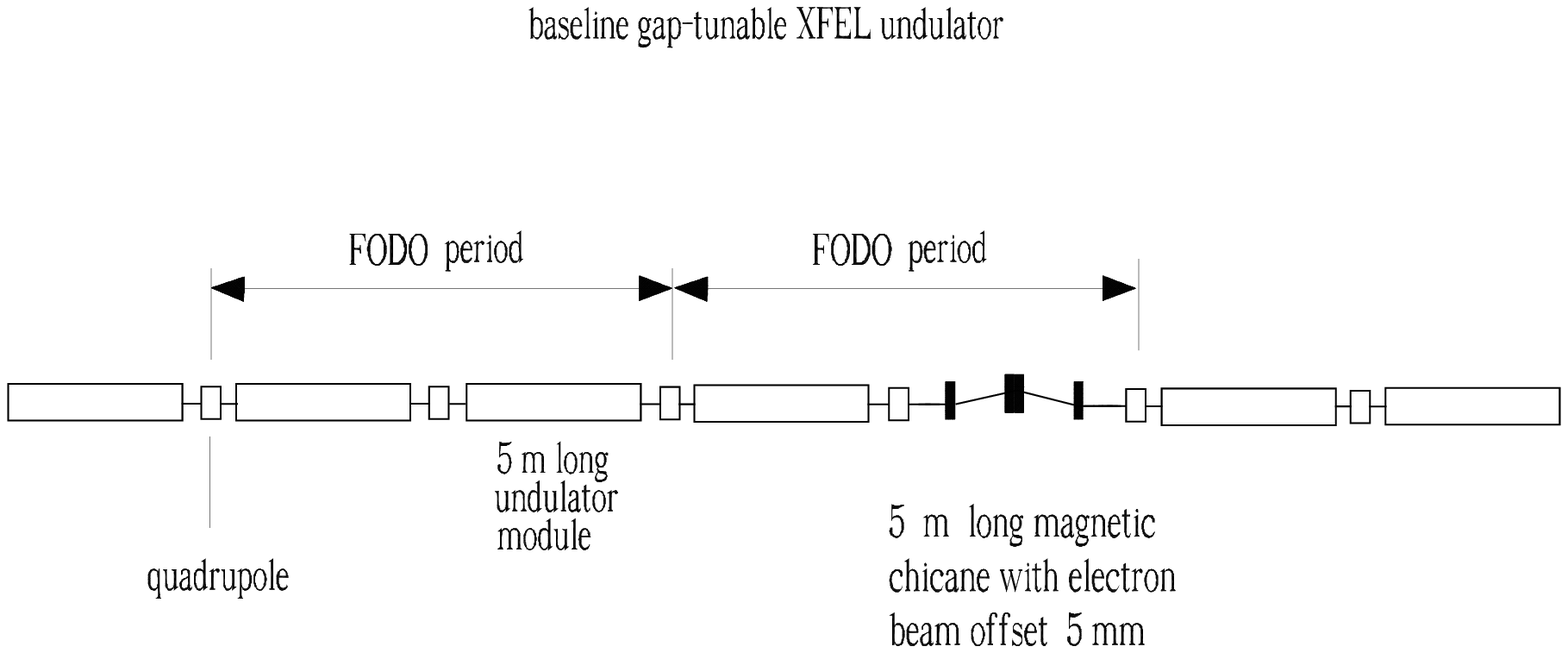}
\caption{Installation of the magnetic chicane in the baseline XFEL
undulator. The magnetic chicane absolves three tasks. First, it
suppresses the electron beam modulation. Second, it allows for the
installation of the single-crystal filter. Third, it performs a
temporal windowing operation by delaying the bunch.} \label{fd2}
\end{figure}
The proposed setup is extremely simple, and is composed of two
simple elements: a crystal, Fig. \ref{fd4}, and a short magnetic
chicane, Fig. \ref{fd2}. The magnetic chicane accomplishes three
tasks. It creates an offset for the crystal installation, it
removes the electron microbunching produced in the first
undulator, and it acts as a delay line for the implementation of
the temporal windowing. Thus, using a single crystal installed
within a short magnetic chicane in the baseline undulator as a
bandstop filter, it is possible to decrease the bandwidth of the
radiation well beyond the XFEL design down to $10^{-5}$. The
installation of the magnetic chicane does not perturb the
undulator focusing system and does not interfere with the baseline
mode of operation.  The scheme can work in combination with a
fresh bunch technique \cite{HUAYU}-\cite{OUR05} both for short
($6$ fs) and long ($60$ fs) pulse mode of operation.

\section{\label{sec2} Principles of the self-seeding technique based on the use of a wake
monochromator}

In this section we illustrate our novel method of
monochromatization, based on the use single crystal monochromator.
As already said, this technique takes advantage of the
transmission geometry, where no extra path-delay for the X-ray
pulse is present. The method consists of a combination of a single
bunch self-seeding scheme, based on the use of a single crystal
monochromator, and of a temporal windowing technique.

\begin{figure}[tb]
\includegraphics[width=1.0\textwidth]{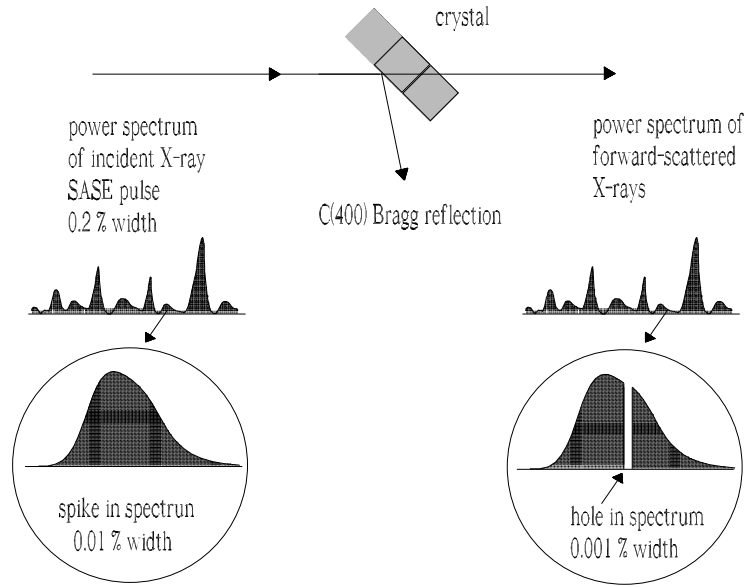}
\caption{Single crystal in Bragg geometry as a bandstop filter for
the transmitted X-ray SASE radiation pulse.} \label{fd7}
\end{figure}

\begin{figure}[tb]
\includegraphics[width=1.0\textwidth]{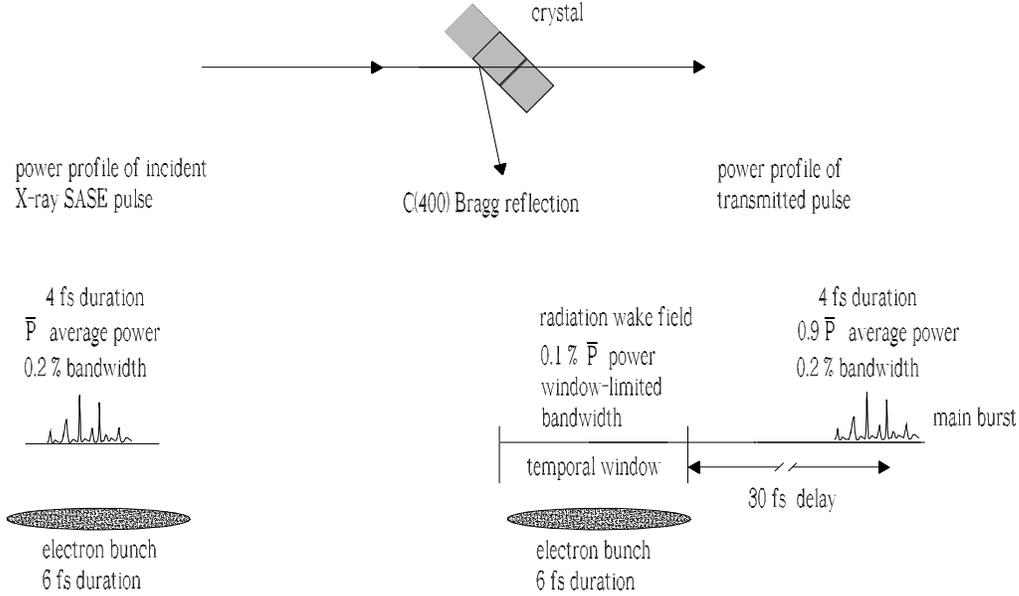}
\caption{Temporal windowing concept. It is possible to eliminate
the spikes in the seed signal by using a temporal window
positioned after the bandstop filter as indicated in the figure.
This can be practically implemented by delaying the electron bunch
at the position where the frequency spectrum of the transmitted
X-ray pulse experiences a strong temporal separation.} \label{fd8}
\end{figure}
The principle of the new method of monochromatization is very
simple and is illustrated in Fig. \ref{fd7} and Fig. \ref{fd8}. An
incident SASE pulse coming from the first undulator impinges on a
crystal set for Bragg diffraction. When X-rays impinge upon a
crystal, forward-scattered X-rays are produced. The phase shift
acquired by the forward-scattered X-rays on the passing through
the crystal depends on the refractive index of the crystal. In
general, the refractive index is slightly less than unity and
complex. The refractive index, however, requires a correction when
the incident beam almost satisfies the Bragg diffraction condition
and multiple scattering takes place. Usually, X-ray multiple
scattering in a perfect crystal is described by the dynamical
theory of X-ray diffraction \cite{AUTH}.  According to this
theory, when the incident angle is near the diffraction condition,
the transmittance spectrum of a thick crystal shows an absorption
line with a narrow width, which is close to the line width in the
reflectance spectrum for the case of small absorption influence.
In this paper we will discuss only the case for small absorption:
in particular, we will consider the C(400) reflection of $0.15$ nm
X-ray from  a $0.1$ mm-thick diamond plate, see Fig. \ref{fd4}.
When the incident angle and the spectral contents of the incoming
beam satisfies the Bragg diffraction condition, the temporal
waveform of the transmitted radiation pulse shows a long
monochromatic wake. The duration of this wake is inversely
proportional to the bandwidth of the absorption line in the
transmittance spectrum. Then, the single crystal in Bragg geometry
actually operates as a bandstop filter for the transmitted X-ray
SASE radiation pulse, see Fig. \ref{fd7}.  Obviously, if we use a
bandstop filter there is no monochromatization in the frequency
domain.  However, it is possible to reach a bandwidth limited seed
signal by using a temporal window positioned after the bandstop
filter as indicated in Fig. \ref{fd8}. In the XFEL case we deal
with a parametric amplifier where the properties of the active
medium, i.e. electron beam, depend on time. As a result, the
temporal windowing concept can be practically implemented in a
simple way by delaying the electron bunch at the position where
the frequency spectrum of the transmitted X-ray pulse experiences
a strong temporal separation. In other words, the magnetic chicane
in Fig. \ref{fd2} shifts the electron bunch on top of the
monochromatic wake created by the bandstop filter. By this, it is
possible to seed the electron bunch with a radiation pulse
characterized by a bandwidth much narrower than the natural FEL
bandwidth. In the proposed scheme, no time dependent elements are
used and problems with synchronization and time jitter do not
exist at all.

\section{\label{sec3} Combination of self-seeding and fresh bunch techniques}

\begin{figure}[tb]
\includegraphics[width=1.0\textwidth]{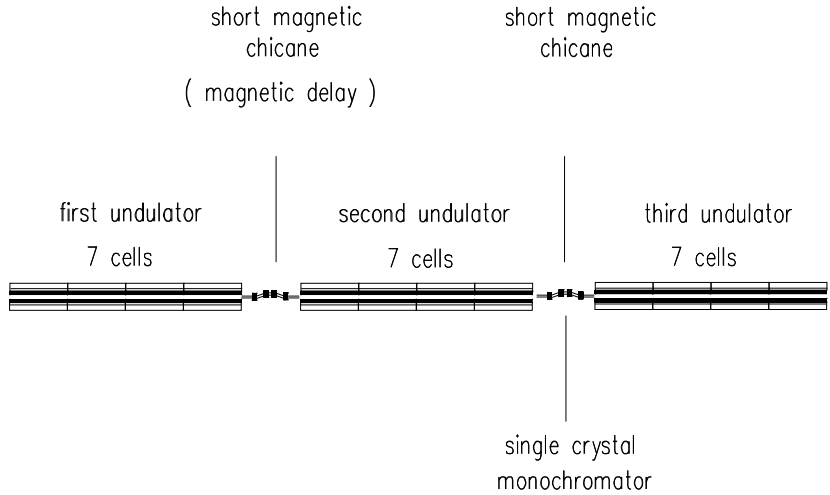}
\caption{Design of an undulator system for the narrow bandwidth
mode of operation. The method exploits a combination of a single
bunch self-seeding scheme, based on the use of a single crystal
monochromator and of a temporal windowing technique, and of a
fresh bunch scheme.} \label{fd1}
\end{figure}
\begin{figure}[tb]
\includegraphics[width=1.0\textwidth]{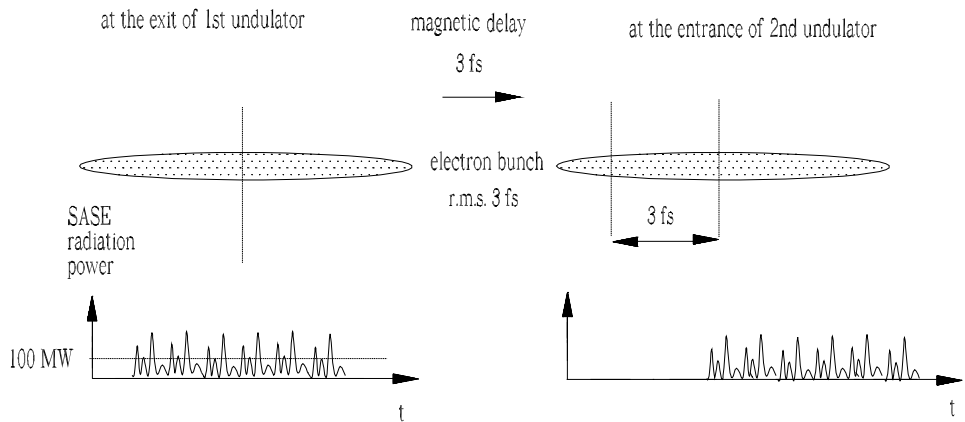}
\caption{Sketch of principle of the fresh bunch technique for the
short ($6$ fs) pulse mode of operation.} \label{fd6}
\end{figure}
\begin{figure}[tb]
\includegraphics[width=1.0\textwidth]{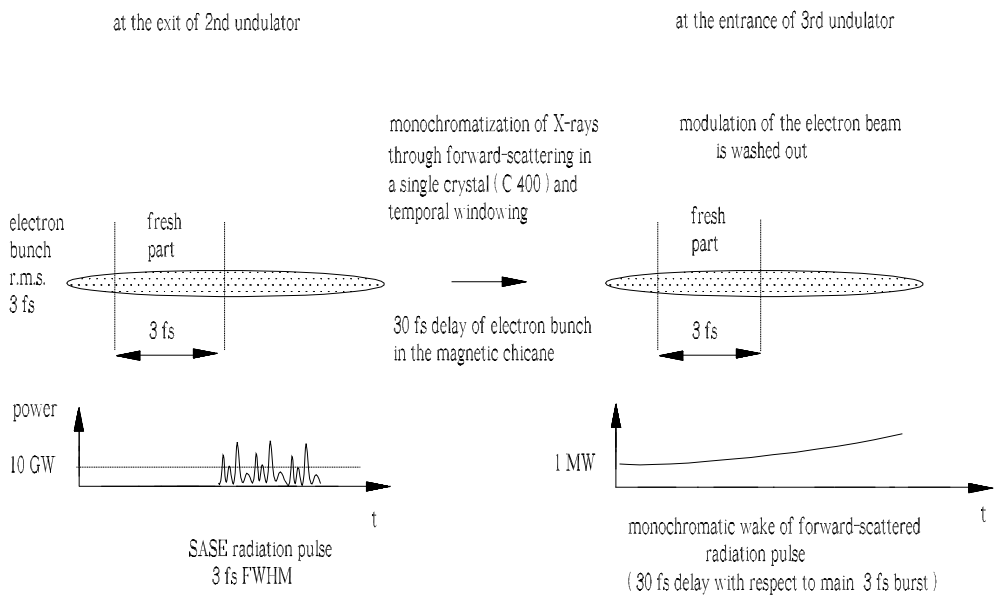}
\caption{Sketch of the X-ray monochromatization principle from the
second to the third undulator. By using a combination of a fresh
bunch technique and a single-bunch self-seeding technique, based
on the use of a single crystal monochromator and of the temporal
windowing operation, a MW-level monochromatic wake of radiation
can be produced behind the monochromator.} \label{fd3}
\end{figure}
In this section we will discuss how the above-described method may
be combined with a fresh bunch technique
\cite{OUR01}-\cite{OUR05}. The idea is sketched in Fig. \ref{fd1},
Fig. \ref{fd6}, and Fig. \ref{fd3}. The scheme can be practically
realized by using three undulator parts, Fig. \ref{fd1}. The first
undulator operates in the linear high-gain regime starting from
the shot noise in the electron beam. After the first undulator,
the electron bunch is sent to a short magnetic chicane, which
removes the microbunching and introduces a delay of the electron
bunch with respect to the radiation pulse, Fig. \ref{fd6}. In this
way, half of the electron bunch is seeded, and saturates in the
second part of the undulator. Note that for European XFEL or LCLS
parameters, microbunching is washed out already with a small
dispersive strength $R_{56}$ in the order of ten microns, allowing
for a short $5$ m-long chicane to be installed in place of a
single undulator module. After the second undulator the electron
bunch is sent to a second magnetic chicane, Fig. \ref{fd3}, and
the strong radiation pulse impinges on a crystal set for Bragg
diffraction. The refracted radiation pulse seeds the remaining
fresh part of the delayed electron bunch, thus implementing the
temporal windowing technique, and is amplified up to saturation in
the third part of the undulator. In this scheme, the mean value of
the radiation power after the crystal within the temporal window
will be in the few MW level. This is the input radiation power at
the entrance of the third undulator, and is much larger than the
shot noise power ($3$ kW).

The combination of self-seeding  and fresh bunch techniques is
extremely insensitive to noise and non-ideal effects. In fact, the
radiation pulse used to produce the monochromatic wake is in the
GW level power. This large power can tremendously improve the
signal to noise ratio of the self-seeding scheme.  It should be
remarked that the possibility of combining self-seeding scheme and
fresh bunch  technique would be especially of great importance
during the early experimental stage, when a proof of principle is
built. During this stage all non ideal effects can be reduced to a
minimum. After this stage, one may implement other self-seeding
schemes based on the wake monochromator.

The advantage of our novel self-seeding method is a minimal
hardware requirement, consisting in the installation of two short
magnetic chicanes as shown in Fig. \ref{fd1}. These chicanes do
not perturb the undulator focusing system and do not interfere
with the baseline mode of operation. Our three stage self-seeding
scheme is therefore compatible with the baseline design of the
European XFEL \cite{tdr-2006}.

\section{\label{sec4} Transmissivity relevant to Bragg diffraction}

Consider the forward-scattered X-rays produced in the transmission
direction when an X-ray beam is directed upon a crystal. As
discussed before, when the incident beam almost satisfies the
Bragg reflection condition, multiple scattering takes place and
the phase shift acquired by the forward-scattered X-ray pulse with
respect to the incident pulse requires corrections to account for
this effect. If we know the modulus of the transmittance of the
crystal $T$ we can find the minimal phase shift of the
forward-scattered X-ray pulse with the help of Kramers-Kronig
relations.

Usually, X-ray multiple scattering in a perfect crystal is
described by the dynamical theory of X-ray diffraction (see e.g.
\cite{AUTH} for a comprehensive treatment). Let us first
illustrate and test the method of phase determination based on the
Kramers-Kronig relations in the case of the reflectance for a
thick crystal. In doing this we will account for the well-known
result of dynamical theory that the modulus of the reflectance has
no zeros in the complex plane and, thus, the minimal phase
solution equals the total phase. According to the dynamical theory
of diffraction, when the incident angle is in the vicinity the
diffraction condition, and one neglects absorption, the
reflectance for a thick crystal is analytically given by

\begin{eqnarray}
Z(\eta)  = \left|\eta \pm \sqrt{\eta^2 - 1}\right| \label{ampli}
\end{eqnarray}
\begin{eqnarray}
\Phi(\eta)  = \mathrm{arg}\left[-\eta + \sqrt{\eta^2 -
1}~\right]~.\label{phase}
\end{eqnarray}
Eq. (\ref{ampli}) and Eq. (\ref{phase}) constitute an analytical
expression for the crystal rocking curve. Here the positive or
negative sign in Eq. (\ref{ampli}) should be taken such that
$Z<1$. In literature $\eta$ is known as the deviation parameter.
When absorption is neglected, such parameter varies from $-1$ to
$+1$ when the reflectivity is $100\%$. Note that in our case of
interest we have an angular divergence of the incident photon beam
about a microradiant, which is much smaller than the Darwin width
of the rocking curve ($10 - 15 \mu$rad.) Therefore, we assume that
all frequencies impinge on the crystal at the same angle. When
this incident angle is in the vicinity of the Bragg diffraction
condition, multiple scattering takes place. However, mirror
reflection takes place for all frequencies. Therefore, the
reflected beam has exactly the same divergence as the incoming
beam. The reflectivity coefficient is described by a reflectivity
curve, which can be extracted from the rocking curve of the
crystal and from the knowledge of the Bragg angle. The incident
angle of the X-ray beam which is interpreted as Bragg angle
determines the position of the reflectivity curve along the
frequency axis. Multiplication of $\eta$ by the bandwidth $\Delta
\omega_{nkl}/\omega$, which depends on the reflection, yields the
reflectivity curve and the phase as a function of the deviation
from the center of reflectivity curve, $\Delta \omega/\omega =
\eta \Delta \omega_{nkl}/\omega$. Thus, in our case we can say
that $\eta$ represents a reduced frequency deviation\footnote{It
may be worth to note here that usually one is interested in the
reflectivity curve, at a fixed frequency of incident radiation, as
a function of the deviation from the Bragg's angle. Here, instead,
we consider the reflectivity at the Bragg's angle corresponding to
a given frequency $\omega$ and we scan the frequency $\Delta
\omega$ at a fixed angle.}, and we can rewrite $Z$ and $\Phi$ as a
function of $\omega$.

$Z(\omega)$ and $\Phi(\omega)$ may thus be regarded as modulus and
phase of a linear filter in the frequency domain. Despite the
numerous achievements of the dynamical theory of diffraction it
has never been considered, at least to our knowledge, that the
phase of this filter, $\Phi(\omega)$, can be recovered from the
knowledge of the modulus $Z(\omega)$ alone just by exploiting
causality and square-integrability, which are both obvious
physical requisites, yielding an interesting relation between
dynamical theory of diffraction and Kramers-Kronig relations.

In fact, according to Titchmarsch theorem (see \cite{LUCC} for a
review on the subject) causality\footnote{Causality simply
requires that the filter can respond to a physical input after the
time of that input and never before.} and square-integrability of
the inverse Fourier transform of $\bar{F}(\omega) = Z \exp(i
\Phi)$, which will be indicated with $F(t)$, is equivalent to the
existence of an analytic continuation of $Z \exp(i \Phi)$ to
$\Omega = \omega + i \omega'$ on the upper complex $\Omega$-plane
(i.e. for $\omega'>0$). The same theorem, also shows that the two
previous statements are equivalent to the fact that real and
imaginary part of $\bar{F}(\omega)$ are connected by Hilbert
transformation. Since $F(t)$ must be real (thus implying that
$\bar{F}^*(\omega)=\bar{F}(-\omega)$), from the Hilbert
transformation follow the well-known Kramers-Kroninig relations
\cite{KRAM,KRON}:

\begin{eqnarray}
&&\mathrm{Re}[\bar{F}(\omega)]=\frac{2}{\pi}\mathcal{P}
\int_0^{\infty}
\frac{\omega'\mathrm{Im}[\bar{F}(\omega')]}{\omega'^2-\omega^2}
d\omega' \cr &&
\mathrm{Im}[\bar{F}(\omega)]=-\frac{2}{\pi}\mathcal{P}
\int_0^{\infty}
\frac{\mathrm{Re}[\bar{F}(\omega')]}{\omega'^2-\omega^2}
d\omega'~,
 \label{KKrel}
\end{eqnarray}
linking real and imaginary part of $\bar{F}(\omega)$. A similar
reasoning can be done for the modulus $Z(\omega)$ and the phase
$\Phi(\omega)$, see \cite{TOLL}. In fact, one can write

\begin{eqnarray}
\mathrm{ln}[\bar{F}(\omega)] = \mathrm{ln}[Z(\omega)] + i
\Phi(\omega)~.\label{ln}
\end{eqnarray}
Note that $\bar{F}^*(\omega)=\bar{F}(-\omega)$ implies that
$|\bar{F}(\omega)|=|\bar{F}(-\omega)|$ and that $\Phi(\omega) = -
\Phi(-\omega)$. Therefore, using Eq. (\ref{ln}) one also has that
$\mathrm{ln}[\bar{F}(\omega)]^*=\mathrm{ln}[\bar{F}(-\omega)]$.
Then, similarly as before, application of Titchmarsh theorem shows
that the analyticity of $\mathrm{ln}[Z(\Omega)]$ on the upper
complex $\Omega$-plane implies that

\begin{eqnarray}
\Phi(\omega)=-\frac{2}{\pi}\mathcal{P} \int_0^{\infty}
\frac{\mathrm{ln}[Z(\omega')] }{\omega'^2-\omega^2} d\omega'~,
\label{KKrel2}
\end{eqnarray}
\begin{figure}[tb]
\includegraphics[width=1.0\textwidth]{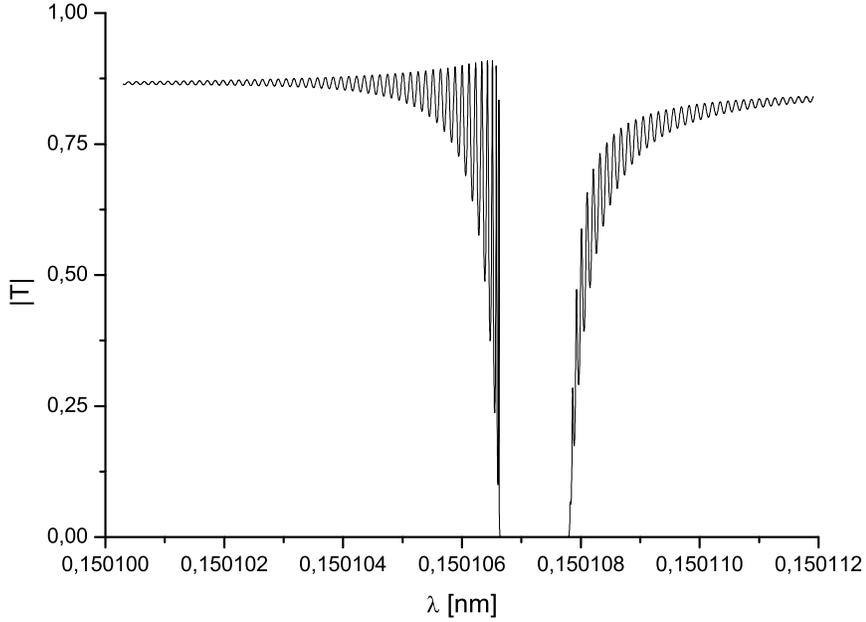}
\caption{Transmissivity (sigma polarization) relevant to the Bragg
400 diffraction of X-rays at $0.15$ nm from a Diamond crystal with
a thickness of $0.1$ mm.} \label{Tmod}
\end{figure}
One may verify that a direct use of Eq. (\ref{KKrel2}), with $Z$
given as in Eq. (\ref{ampli}), yields back the phase
$\Phi(\omega)$. It is important to note, however, that in applying
such procedure we tacitly assumed that $\mathrm{ln}[Z(\Omega)]$ is
analytical on the upper complex $\Omega$-plane. While causality
implies this fact for $\bar{F}(\Omega)$, this is not the case,in
general, for $\mathrm{ln}[Z(\Omega)]$. In fact, such function is
singular where $\bar{F}(\Omega)=0$. In the case under discussion,
one can retrieve $\Phi(\omega)$ only because $\bar{F}(\Omega)>0$
everywhere on the complex plane, as it can be seen taking the
analytic continuation of $\bar{F}(\omega)$ defined by Eq.
(\ref{ampli}) and Eq. (\ref{phase}). Note, however, that the
knowledge of this fact comes as a consequence of diffraction
theory. Without such information we would be able to go through
the same algorithm, but we would not be certain that Eq.
(\ref{KKrel2}) really yields back $\Phi(\omega)$. If
$\bar{F}(\omega)$ had zeros on the upper complex plane, these
zeros would have contributed adding extra terms to the total
phase. For this reason, Eq. (\ref{KKrel2}) is known as minimal
phase solution.

\begin{figure}[tb]
\includegraphics[width=1.0\textwidth]{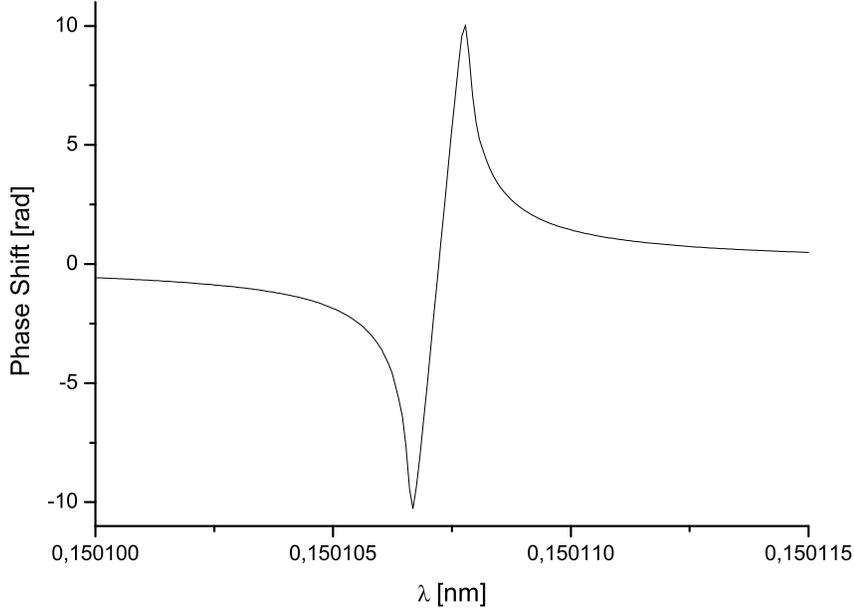}
\caption{Minimal phase shift of the forward-diffracted X-rays at
$0.15$ nm (sigma polarization) relevant to the Bragg 400
diffraction from a Diamond crystal with a thickness of $0.1$ mm.
The phase shift is reconstructed by Kramers-Kronig transformation
Eq (\ref{KKrel2}). } \label{Targ}
\end{figure}

Let us now consider the phase shift of the forward-diffracted
X-rays by a crystal with absorption. We first used the dynamical
theory of diffraction to evaluate the transmittance in Bragg
geometry. Fig. \ref{Tmod} gives the transmissivity curve for a
thick ($0.1$ mm) absorbing crystal in Bragg geometry for the
Diamond 400 reflection \cite{WECK}. We subsequently used the
modulus of the transmittance in Fig. \ref{Tmod}, and we used the
software in \cite{LUC2}, accompanying reference \cite{LUCC}, to
retrieve the minimal phase solution according to Eq.
(\ref{KKrel2}). The result, which will be used in the following
sections, is shown in Fig. \ref{Targ}.

Finally, it is important to remark that the reconstructed phase
shift satisfies causality, meaning that we found a causal solution
for the transmitted X-rays. Without taking into account the phase
shift of our transmitted pulse, causality would be obviously
violated. We combined dynamical theory and Kramers-Kronig
relations for the first time to our knowledge. This combination
may also be used for other purposes, for example to check
numerical calculations based on the dynamical theory of X-ray
diffraction. One may now check the shape of real and imaginary
parts of reflectance and transmittance in relation with causality.

\section{\label{sec5} Feasibility study}

Following the introduction of the proposed methods we report on a
feasibility study of the single-bunch self-seeding scheme with a
wake monochromator. This feasibility study is performed with the
help of the FEL code GENESIS 1.3 \cite{GENE} running on a parallel
machine. In the next subsection we will present the feasibility
study for the short-pulse mode of operation ($6$ fs), while, later
on, we will cover the long-pulse mode of operation ($60$ fs).
Parameters used in simulations for the short pulse mode of
operations are presented in Table \ref{tt1}. For the long pulse
mode of operations Table \ref{tt1} is still valid, except for a
ten times larger charge ($0.25$ nC) and a ten times longer rms
bunch length ($10 \mu$m, corresponding to $60$ fs). In both cases
we use a combination of the fresh bunch technique and the wake
monochromator method, shown in Fig. \ref{fd1}. Details on the
operation of the fresh-bunch technique and its combination with
the self-seeding technique are discussed in \cite{OURL}, and the
first two undulator parts considered in that reference are exactly
the same used here.

\begin{table}
\caption{Parameters for the short pulse mode of operation used in
this paper.}

\begin{small}\begin{tabular}{ l c c}
\hline & ~ Units &  ~ \\ \hline
Undulator period      & mm                  & 48     \\
K parameter (rms)     & -                   & 2.516  \\
Wavelength            & nm                  & 0.15   \\
Energy                & GeV                 & 17.5   \\
Charge                & nC                  & 0.025 \\
Bunch length (rms)    & $\mu$m              & 1.0    \\
Normalized emittance  & mm~mrad             & 0.4    \\
Energy spread         & MeV                 & 1.5    \\
\hline
\end{tabular}\end{small}
\label{tt1}
\end{table}

\subsection{\label{sec5A} Feasibility study for the short pulse mode of
operation}

\begin{figure}[tb]
\includegraphics[width=1.0\textwidth]{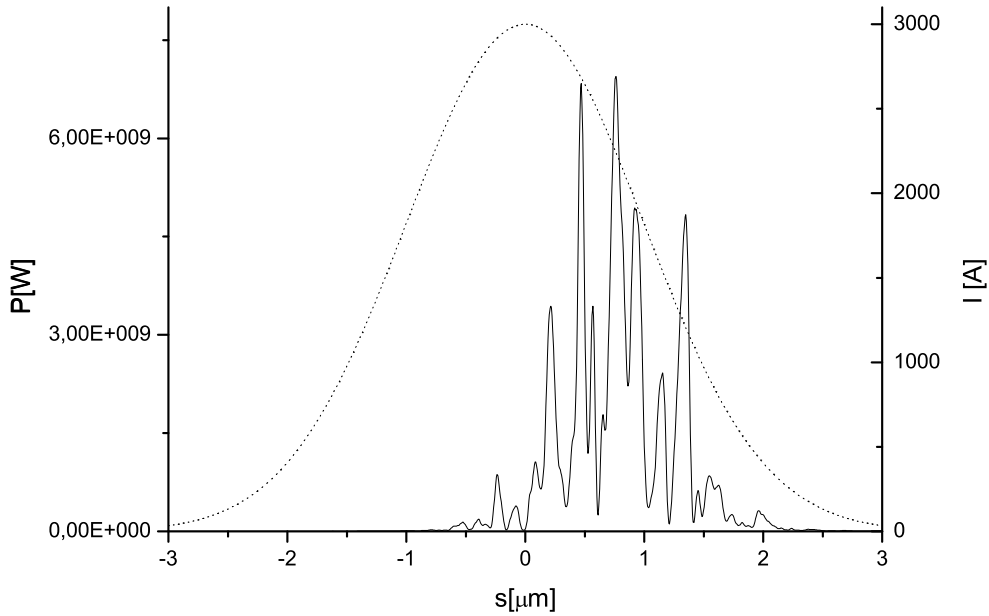}
\caption{Short pulse mode of operation, combination of
self-seeding and fresh bunch techniques. Output power at the end
of the second stage, $7$ cells long ($42$ m). The dashed line
illustrates the corresponding distribution of the electron beam
current.} \label{Pinp}
\end{figure}

We begin with Fig. \ref{Pinp}, where we present the output at the
end of the second stage, i.e. after the application of the
fresh-bunch technique. We take this result as our starting point
without further comments, since Fig. \ref{Pinp} has already been
introduced and thoroughly discussed in \cite{OURL}.

\begin{figure}[tb]
\includegraphics[width=1.0\textwidth]{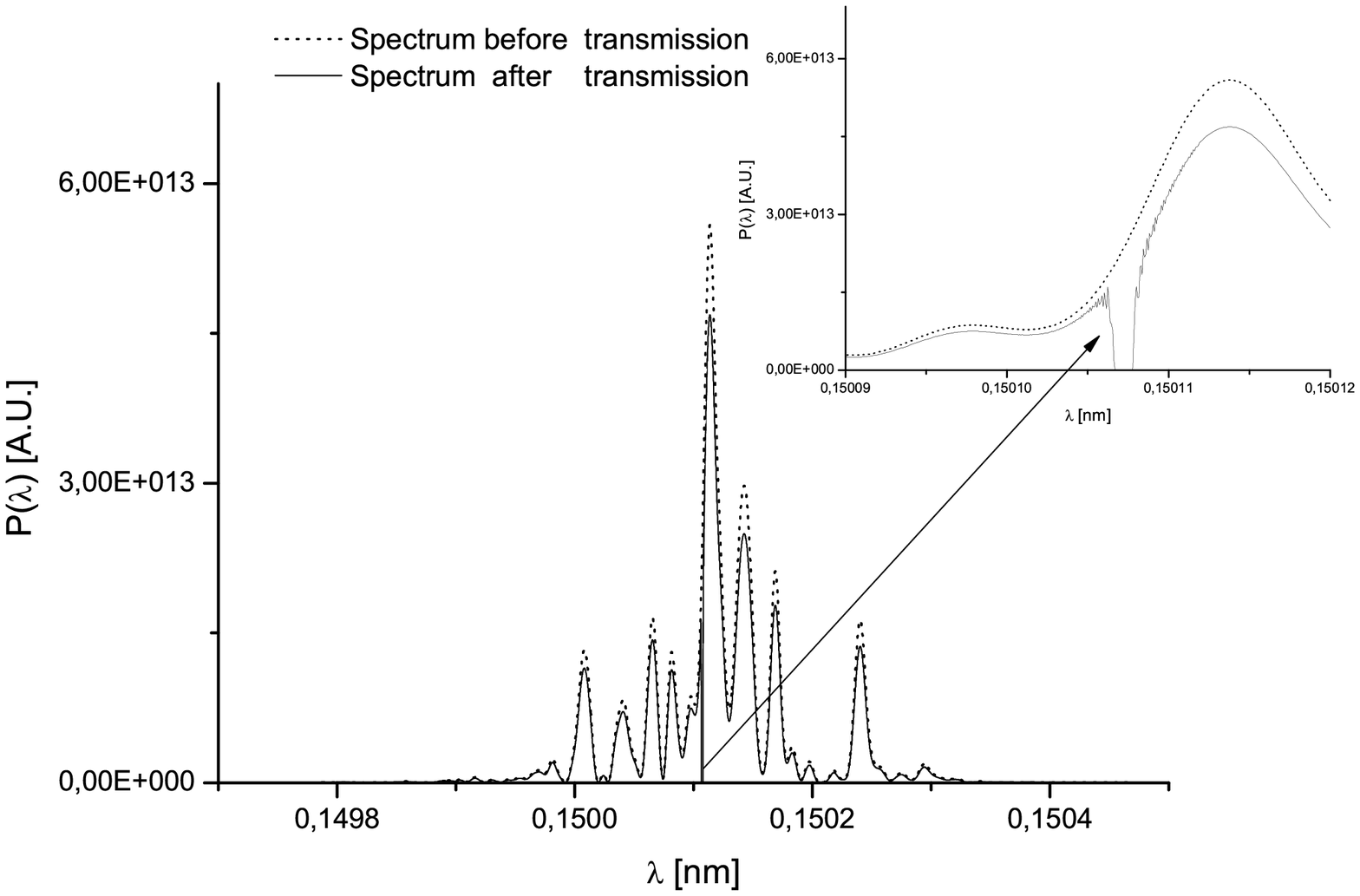}
\caption{Short pulse mode of operation, combination of
self-seeding and fresh bunch techniques. Output spectrum after the
diamond crystal. The bandstop effect is clearly visible, and
highlighted in the inset. For comparison, the spectrum before
transmission (dotted line) is superimposed to the spectrum after
transmission (solid line).} \label{Spinout}
\end{figure}
\begin{figure}[tb]
\includegraphics[width=1.0\textwidth]{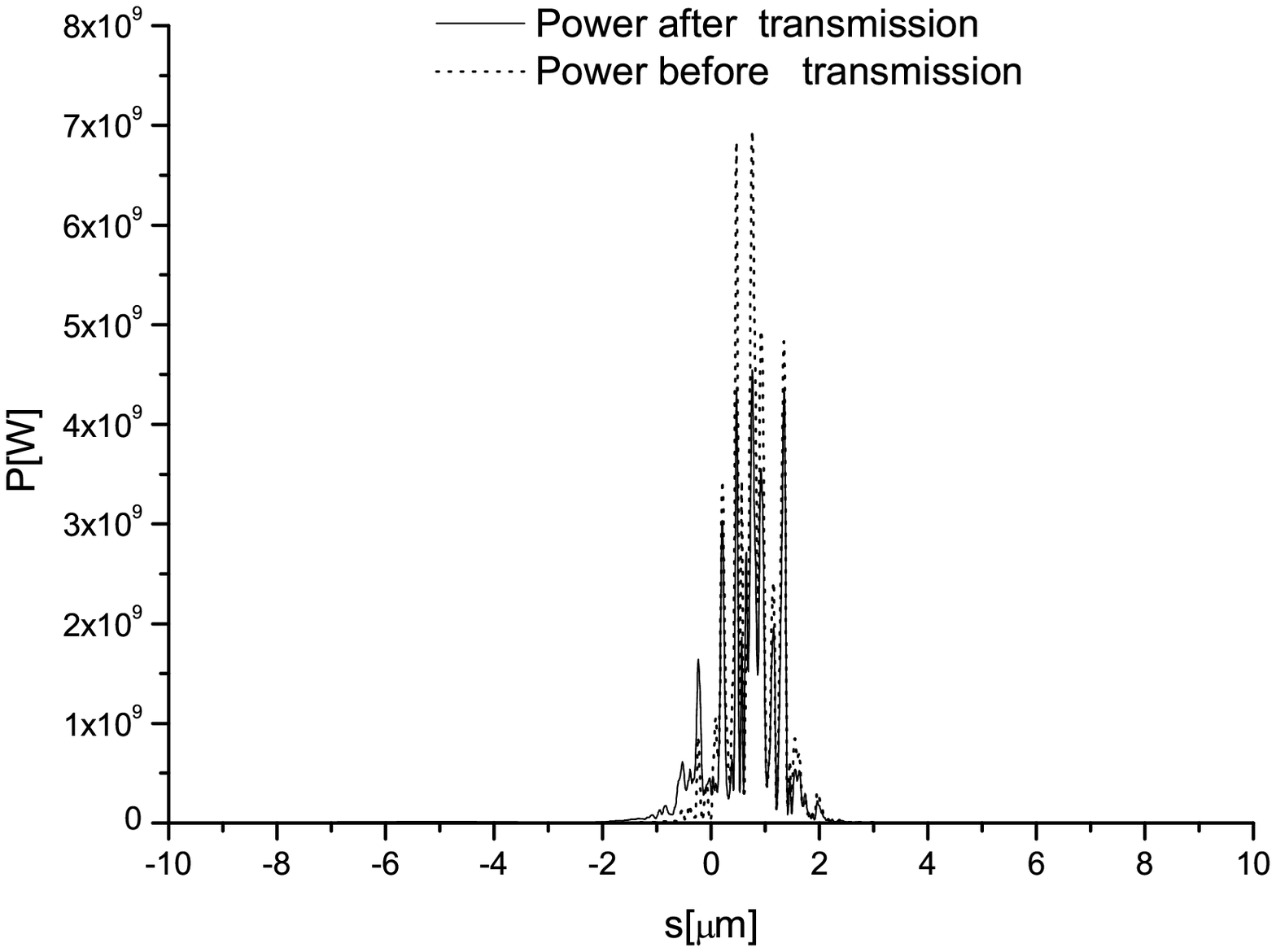}
\caption{Short pulse mode of operation, combination of
self-seeding and fresh bunch techniques. Power distribution before
(dotted line) and after (solid line) transmission through the
crystal.} \label{Pout}
\end{figure}
\begin{figure}[tb]
\includegraphics[width=1.0\textwidth]{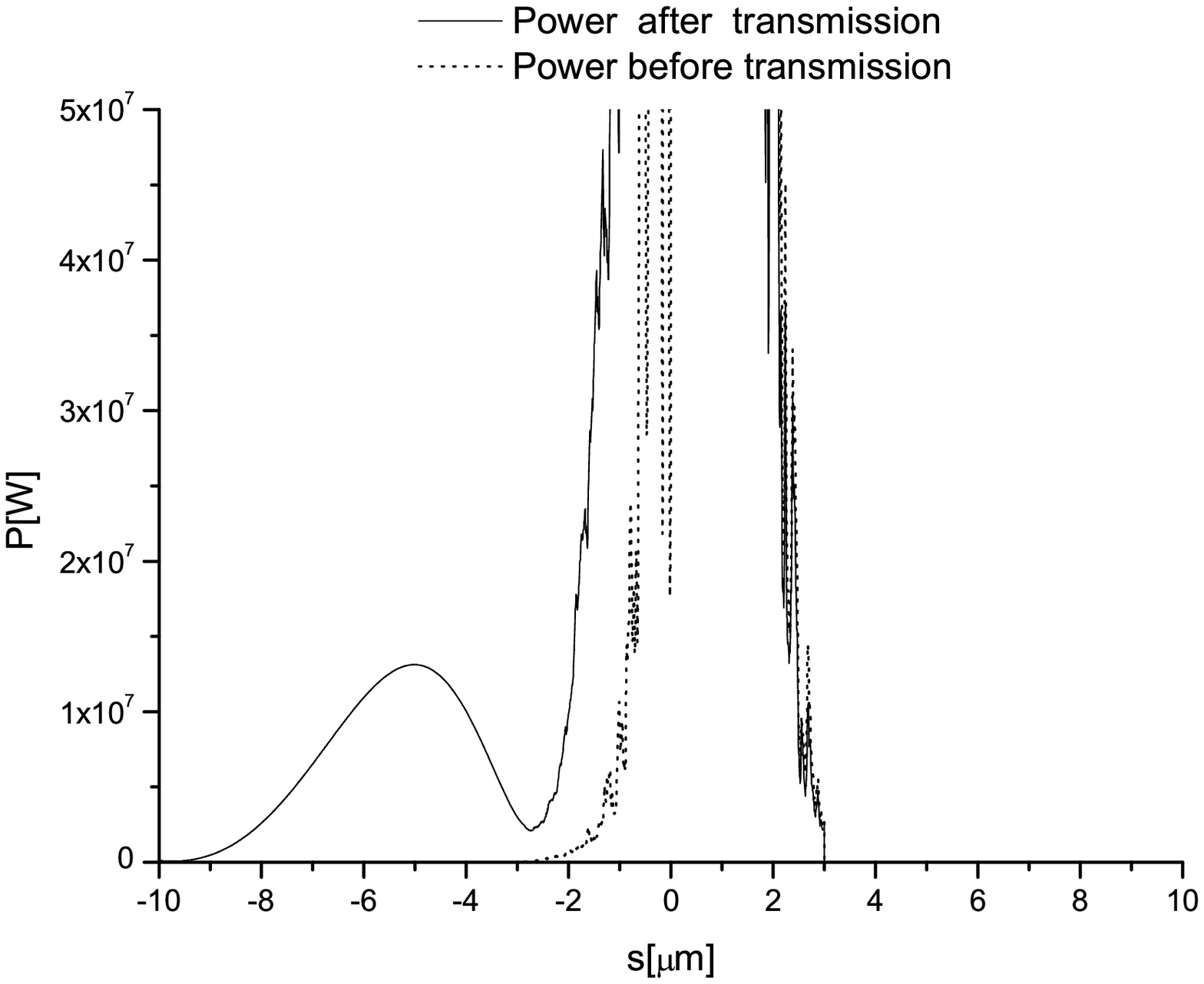}
\caption{Enlargement of Fig. \ref{Pout}. The horizontal axis is
left unchanged, while the vertical axis is zoomed. The
monochromatic tail due to the transmission through the bandstop
filter is now evident on the left of the figure.} \label{Pouttail}
\end{figure}
\begin{figure}[tb]
\includegraphics[width=1.0\textwidth]{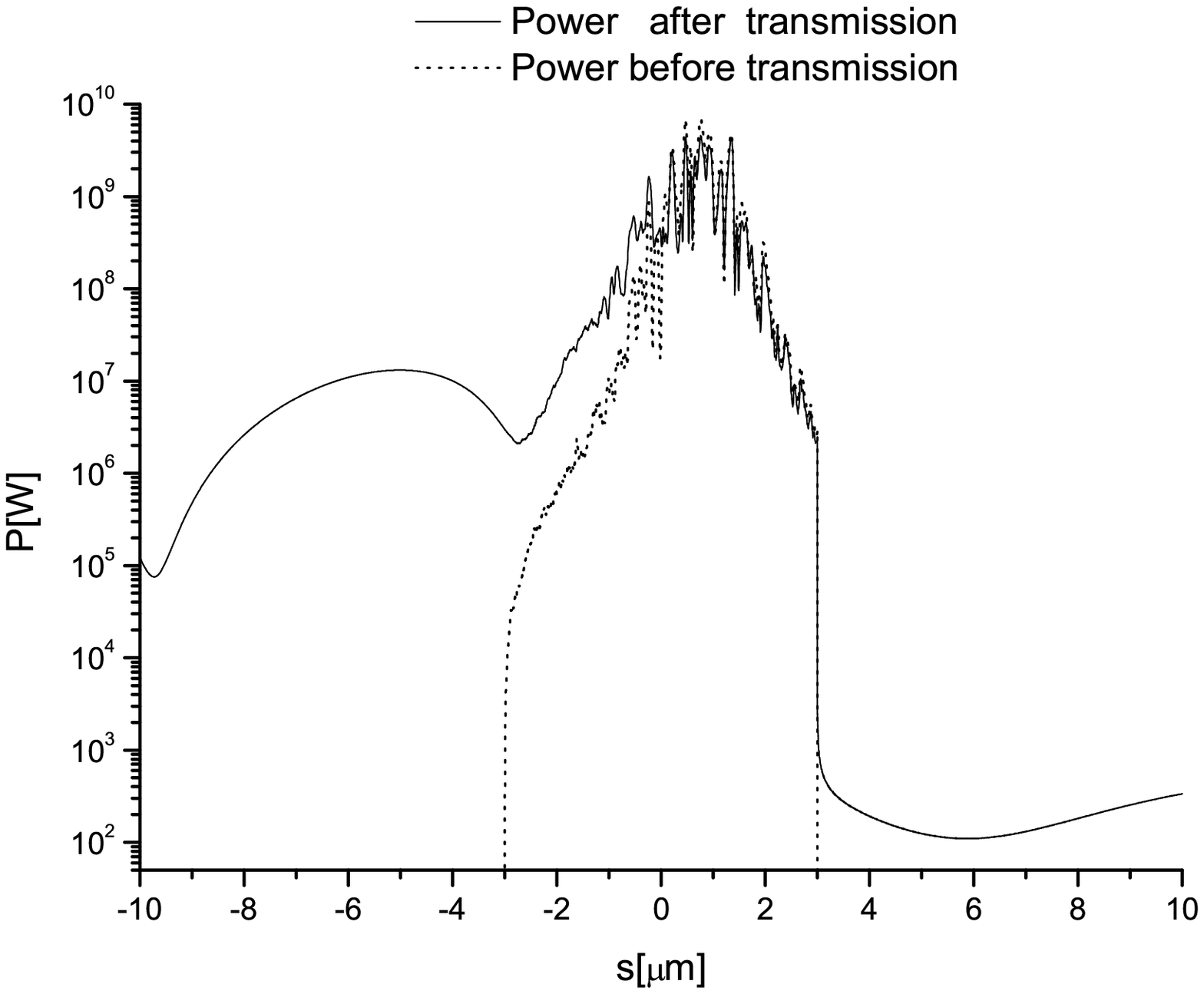}
\caption{Short pulse mode of operation, combination of
self-seeding and fresh bunch techniques. Output power after the
diamond crystal in logarithmic scale. Power distribution after
transmission through the crystal, accounting for the transmission
phase.} \label{Poutlog}
\end{figure}
\begin{figure}[tb]
\includegraphics[width=1.0\textwidth]{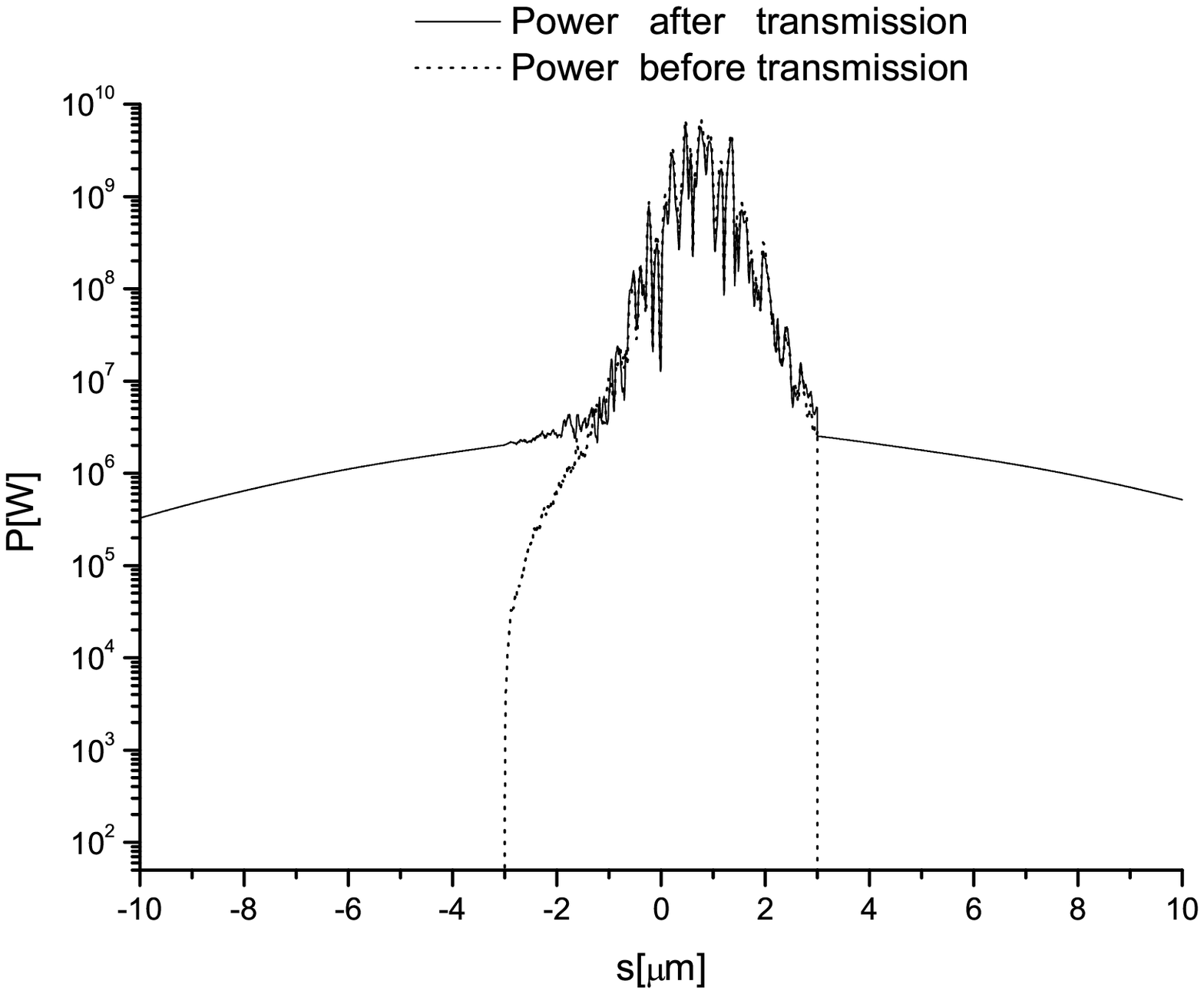}
\caption{Short pulse mode of operation, combination of
self-seeding and fresh bunch techniques. Output power after the
diamond crystal in logarithmic scale. Power distribution after
transmission through the crystal, not accounting for the
transmission phase.} \label{Poutlognoph}
\end{figure}

As explained before, the crystal acts as a bandstop filter. Such
effect is best shown in terms of the spectrum in Fig.
\ref{Spinout}, where we show a comparison between spectra before
and after the filter. The effect is highlighted in the inset. The
corresponding power is shown in Fig. \ref{Pout}. As discussed
before, monochromatization does not take place in the frequency
domain. At first glance, the passage through the bandstop filter
is only responsible for slight changes in the power distribution
along the pulse. However, a zoom of the vertical axis shows what
we are interested in: a long, monochromatic tail in the power
distribution on the left side of the picture, Fig. \ref{Pouttail}.
Note that there is no corresponding tail on the right side of Fig.
\ref{Pouttail}. This fact is consistent with causality. This fact
is evident from the analysis of Fig. \ref{Poutlog}, which
illustrates the power distribution before and after the filter in
logarithmic scale. It is interesting to compare Fig. \ref{Poutlog}
with what would have been obtained in the case we completely
neglected the phase in the transmittance $T$. In this case,
instead of Fig. \ref{Poutlog} we would have obtained Fig.
\ref{Poutlognoph}. It is interesting to see that even without
phase we can rely on sufficient power for seeding. However, we
would have strongly underestimated it of about a factor $10$.
Moreover, as expected, Fig. \ref{Poutlognoph} shows a non-causal
behavior of the signal. This is evident from the analysis of Fig.
\ref{Poutlog}, where we compare the power after transmission with
and without phase, and the power before transmission in
logarithmic scale. The large jump between the power after
transmission accounting for the phase is due to causality. Data
without phase show, instead, a symmetric behavior, which is not
causal at all. The  very small ($\sim 10^{-4}$) but visible
departures from the exact zero result appear in the right-hand
side of the picture in logarithmic scale, Fig. \ref{Poutlog}. This
accuracy is acceptable for most purposes.

\begin{figure}[tb]
\includegraphics[width=1.0\textwidth]{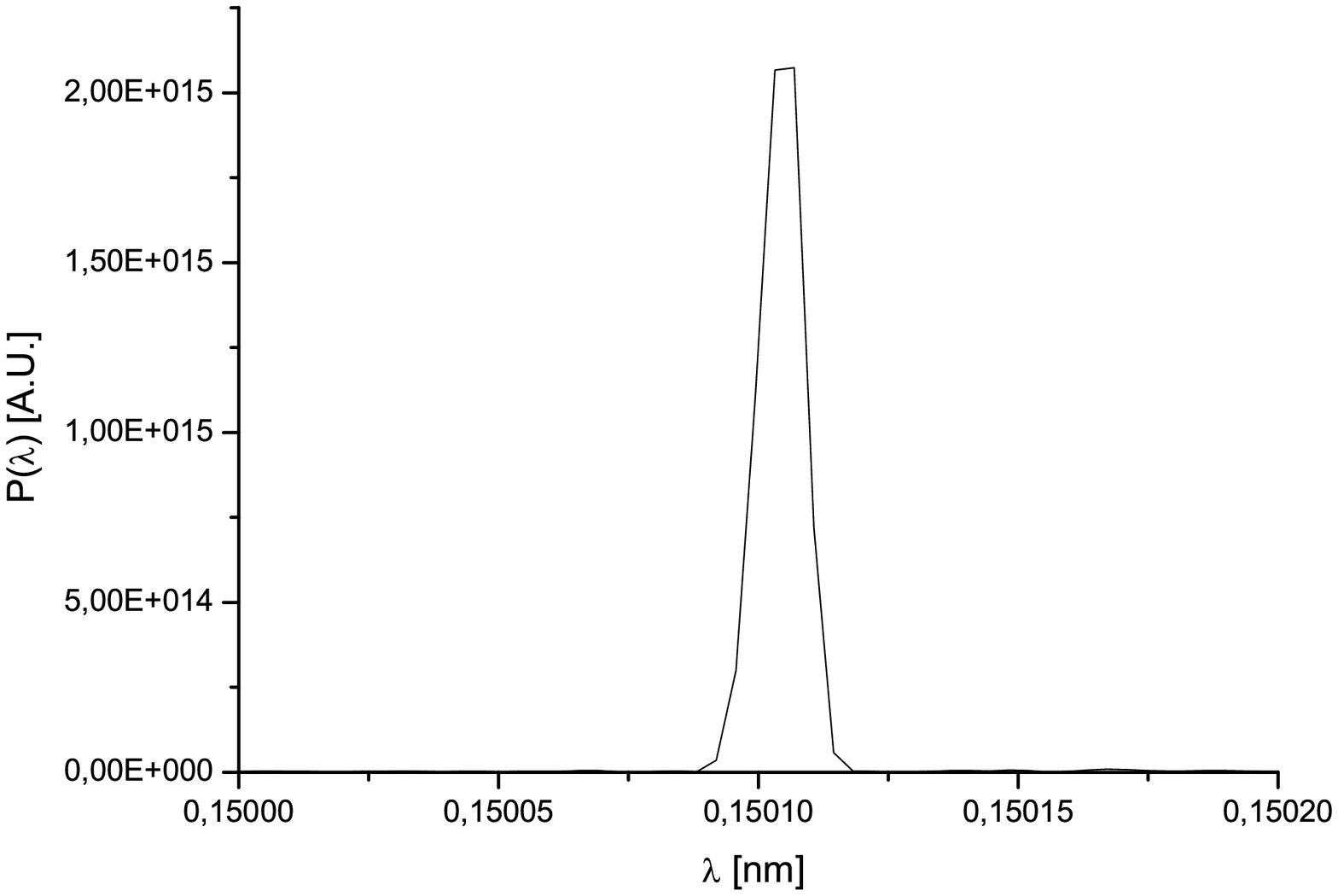}
\caption{Short pulse mode of operation, combination of
self-seeding and fresh bunch techniques. Output spectrum of the
device. The relative spectral width is $5 \cdot 10^{-5}$  and is
close to the transform limit.} \label{Spfinal}
\end{figure}

\begin{figure}[tb]
\includegraphics[width=1.0\textwidth]{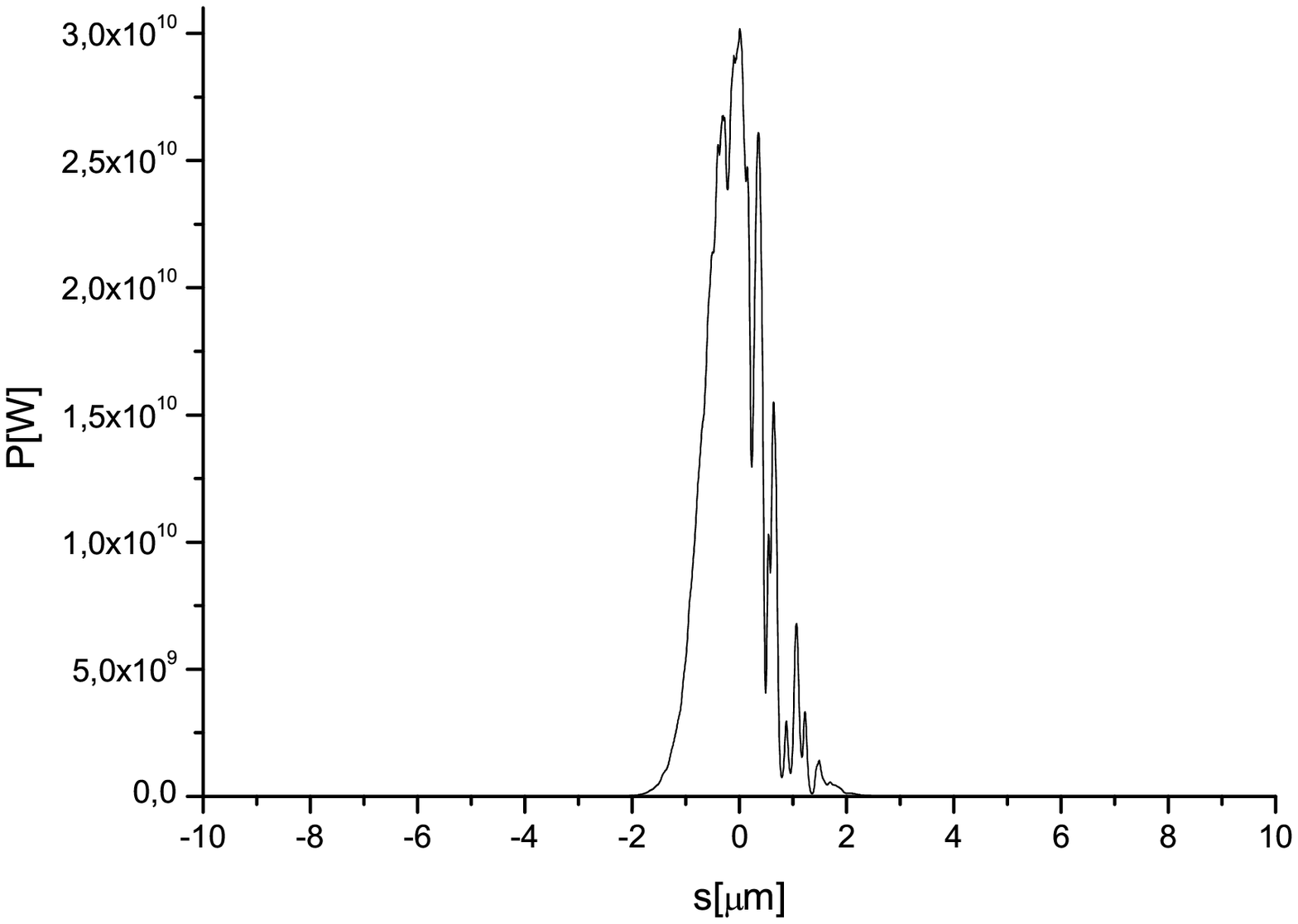}
\caption{Short pulse mode of operation, combination of
self-seeding and fresh bunch techniques. Output power of the
device.} \label{Pfinal}
\end{figure}

Inspection of Fig. \ref{Pouttail} shows that the seeding power in
the monochromatic tail is about $10$ MW. It should be noted that
this power is about three orders of magnitude larger than the shot
noise power. Moreover, within the temporal window that used ($3$
fs, corresponding to about $1 \mu$m in the plot), the seed power
is practically constant along the bunch and fluctuates according
to a negative exponential function i.e. as the instantaneous power
(note that even for $100\%$ fluctuations we always have the seed
power much larger than the shot noise power). As a result, we do
not need to average over many shots, when considering our
feasibility study. It follows that the combination of our new
self-seeding technique with the fresh bunch technique is
particularly well-suited for demonstrating the feasibility of our
technique in a simple but sure way. The fresh part of the bunch is
seeded by such monochromatic tail, and the monochromatic signal is
amplified up to saturation in the last undulator part, which is 7
cells long. The final output is shown in Fig. \ref{Spfinal} and
Fig. \ref{Pfinal}, which show both spectrum and power distribution
after the third undulator stage. The relative spectral width is $8
\cdot 10^{-5}$  and is near to the transform-limited bandwidth of
a Gaussian pulse with the same FWHM.

In closing, it should be noted that the feasibility of the method
is not really sensitive on the phase of the transmittance. In
other words, even for the (non-physical) choice of zero phase, one
would obtain a sufficient seed power to demonstrate the
feasibility of the method.

\subsection{\label{sec5B} Feasibility study for the long pulse mode
of operation}

\begin{figure}[tb]
\includegraphics[width=1.0\textwidth]{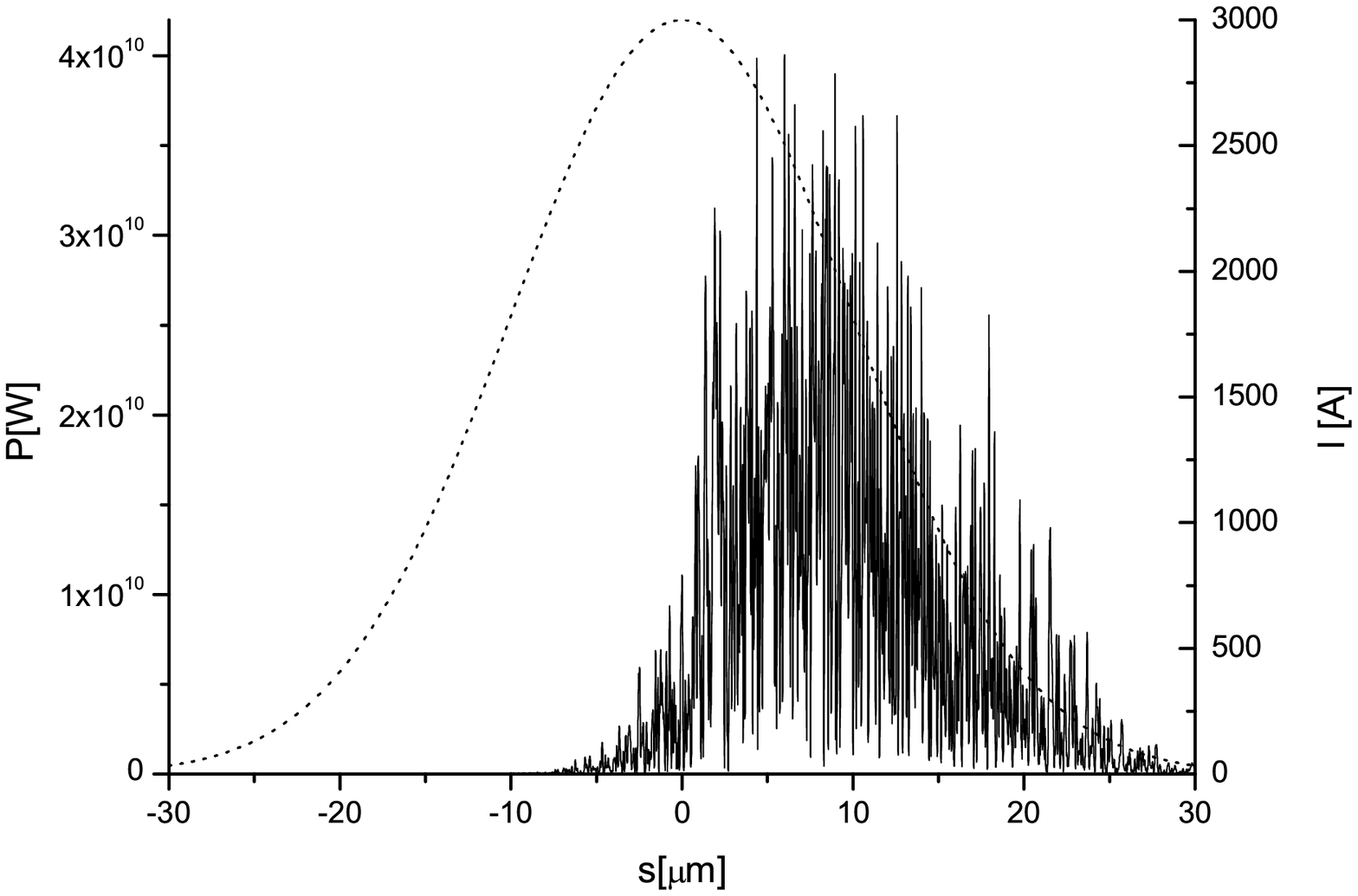}
\caption{Long pulse mode of operation, combination of self-seeding
and fresh bunch techniques. Output power at the end of the second
stage, $7$ cells long ($42$ m). The dashed line illustrates the
corresponding distribution of the electron beam current.}
\label{L2}
\end{figure}
\begin{figure}[tb]
\includegraphics[width=1.0\textwidth]{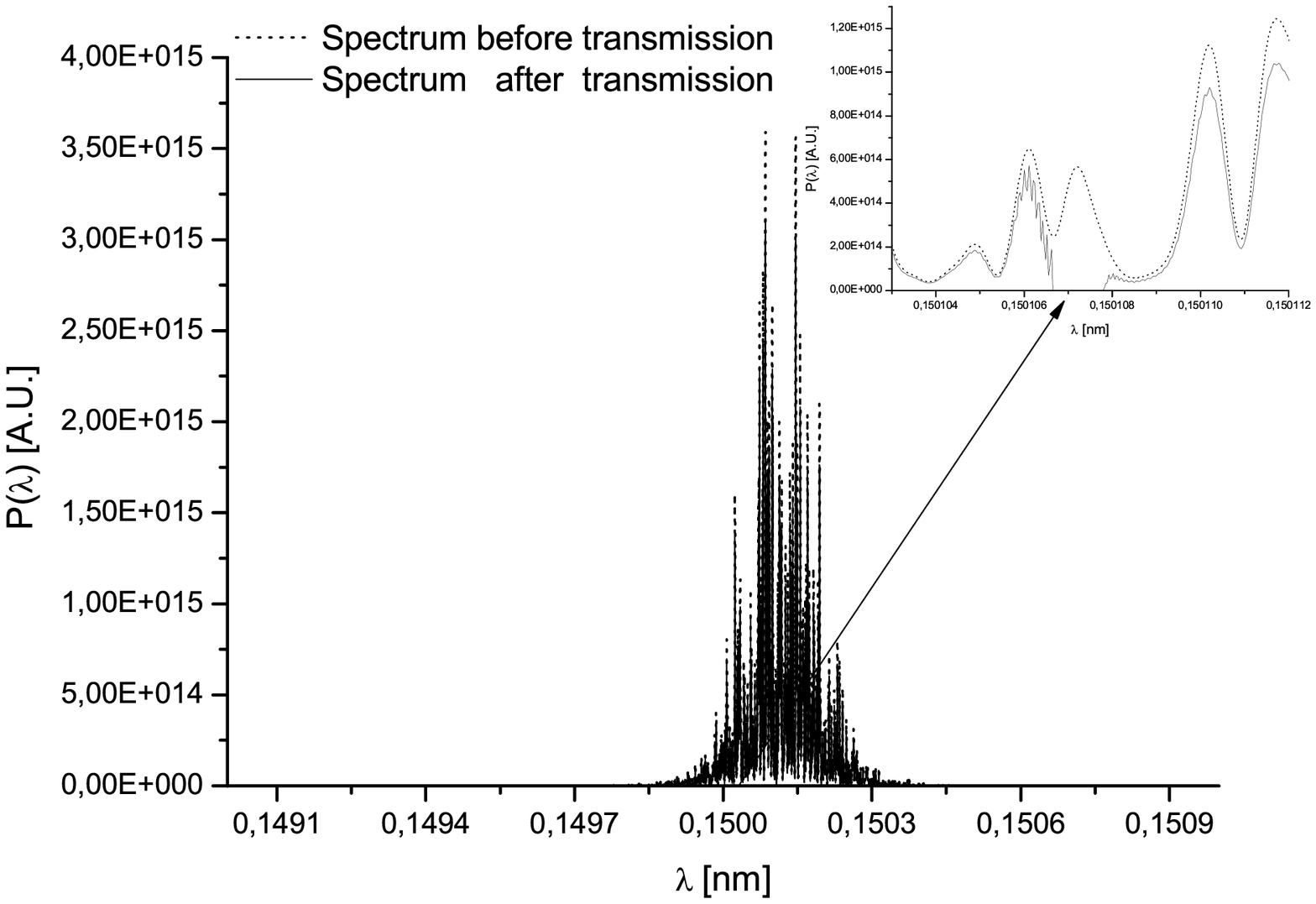}
\caption{Long pulse mode of operation, combination of self-seeding
and fresh bunch techniques. Output spectrum after the diamond
crystal. The bandstop effect is clearly visible, and highlighted
in the inset. For comparison, the spectrum before transmission
(dotted line) is superimposed to the spectrum after transmission
(solid line).} \label{L3}
\end{figure}
\begin{figure}[tb]
\includegraphics[width=1.0\textwidth]{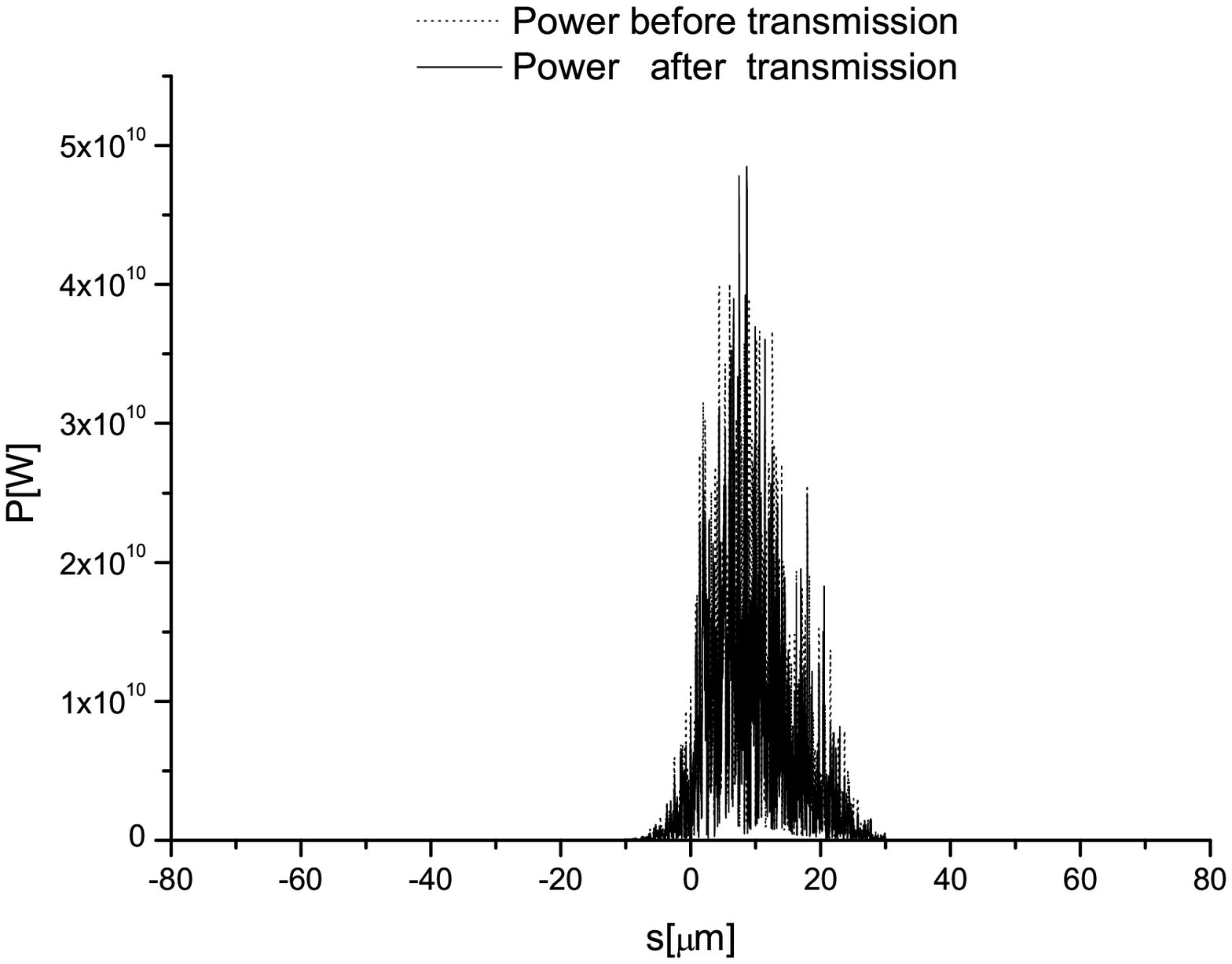}
\caption{Long pulse mode of operation, combination of self-seeding
and fresh bunch techniques. Power distribution after transmission
through the crystal.} \label{Lpout}
\end{figure}
\begin{figure}[tb]
\includegraphics[width=1.0\textwidth]{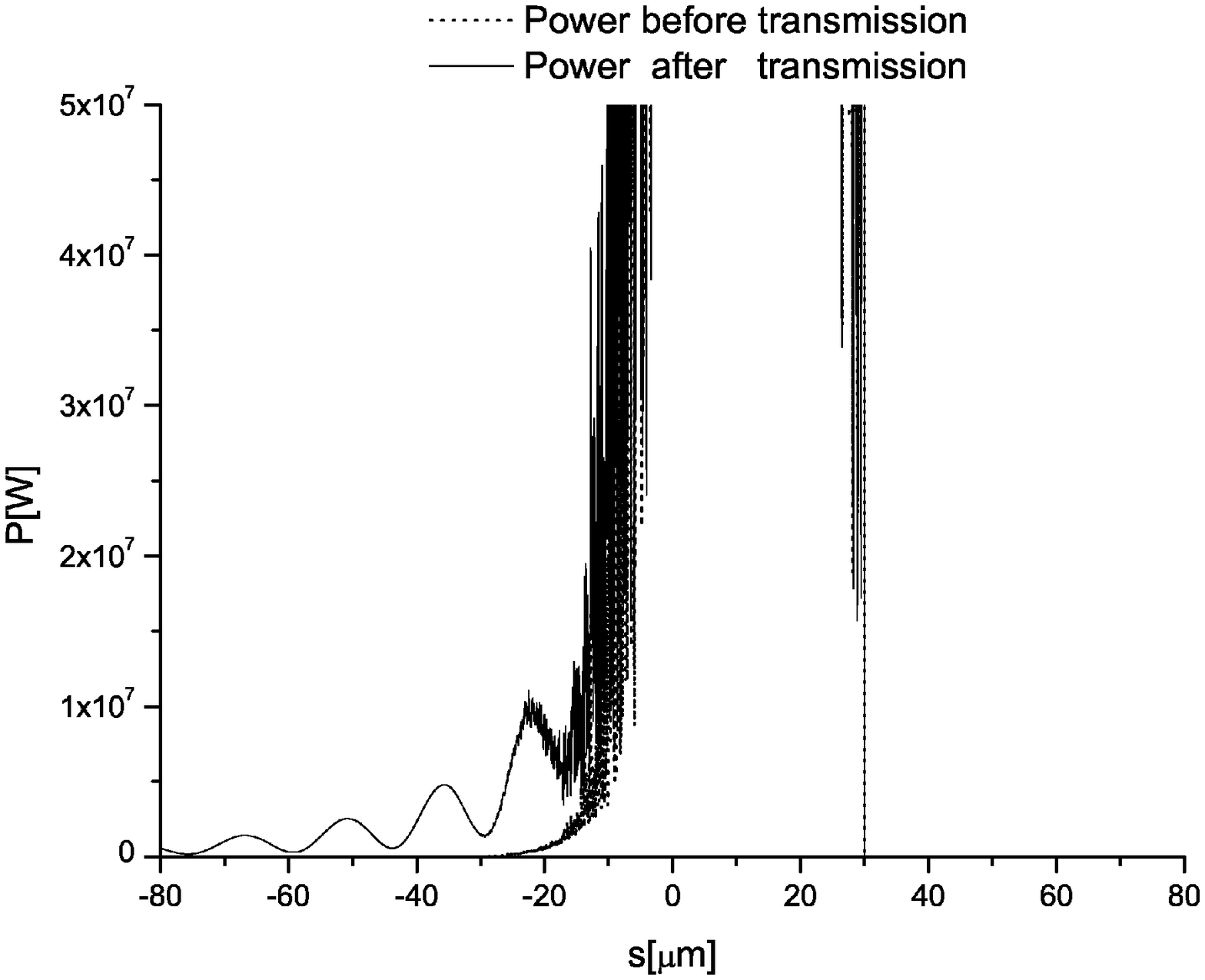}
\caption{Enlargement of Fig. \ref{Lpout}. The horizontal axis is
left unchanged, while the vertical axis is zoomed. The
monochromatic tail due to the transmission through the bandstop
filter is now evident on the left of the figure.}
\label{Lpout_zoom}
\end{figure}

Let us now consider the case of long pulse mode of operation.
Similarly as before, in Fig. \ref{L2} we present the input power.
Subsequently, a comparison between the spectrum before and after
transmission is shown in Fig. \ref{L3}. The bandstop effect is
clearly visible in the inset. In this case we still used the
C(400) reflection. Note that the width of the reflectivity curve
is now comparable with the spike width in the frequency domain.
Thus, in contrast to the short-pulse case, the characteristic
scale of the tail length, which is the inverse line width of the
filter, is comparable with the radiation pulse duration and with
the electron bunch duration, which is even two times longer.
Therefore, it is not trivial to understand how the self-seeding
process takes place. The answer is found by inspection of Fig.
\ref{L3}. The reflectivity curve is, in fact, very sharp, compared
with the spectrum distribution shape. As a result the
monochromatic tail is much longer than the inverse width of the
reflectivity curve. The power distribution is shown in Fig.
\ref{Lpout}, and the monochromatic tail is well-visible in the
zoom of that plot, Fig. \ref{Lpout_zoom}.

\begin{figure}[tb]
\includegraphics[width=1.0\textwidth]{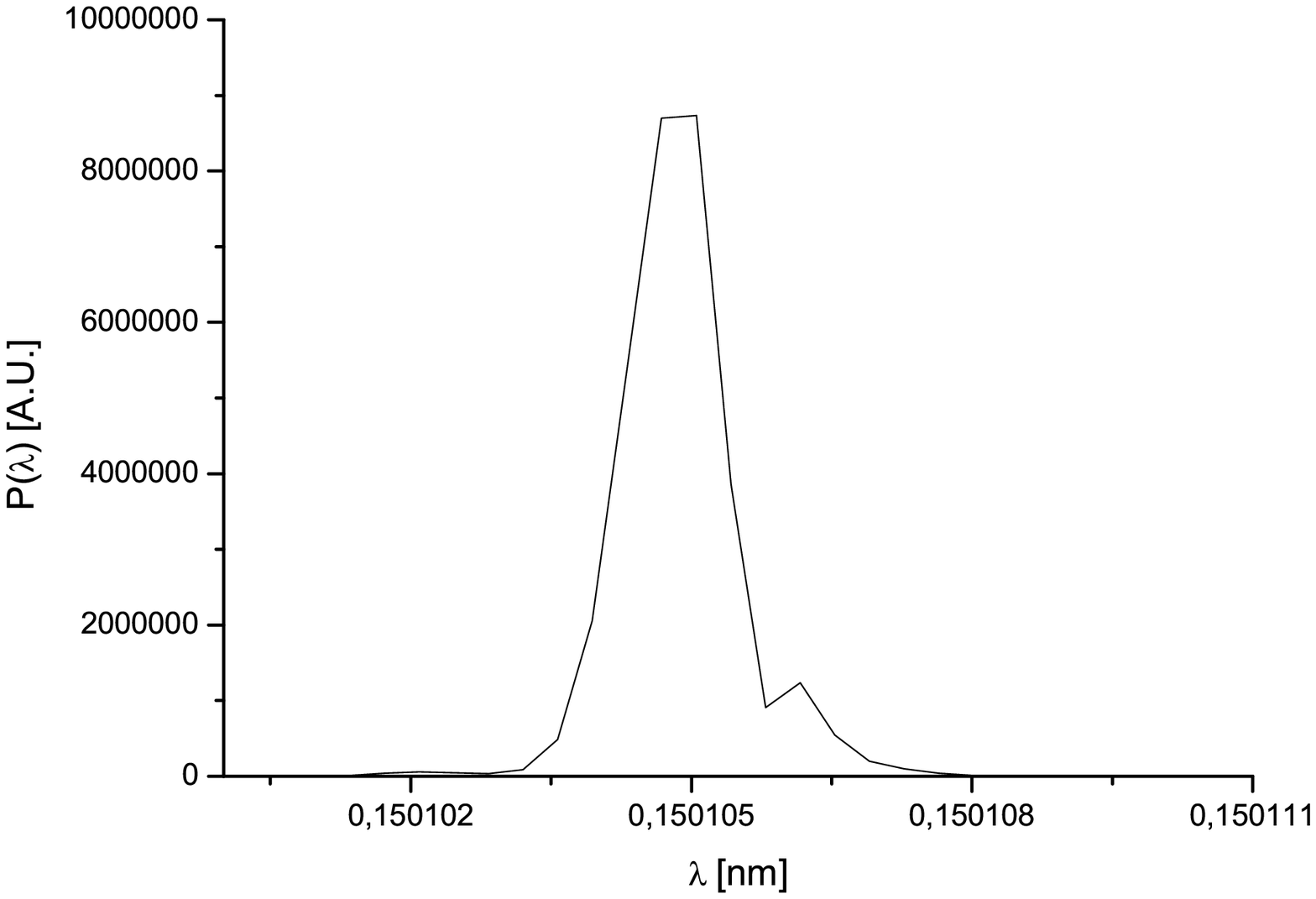}
\caption{Long pulse mode of operation, combination of self-seeding
and fresh bunch techniques. Output spectrum of the device. }
\label{LPfinout}
\end{figure}

\begin{figure}[tb]
\includegraphics[width=1.0\textwidth]{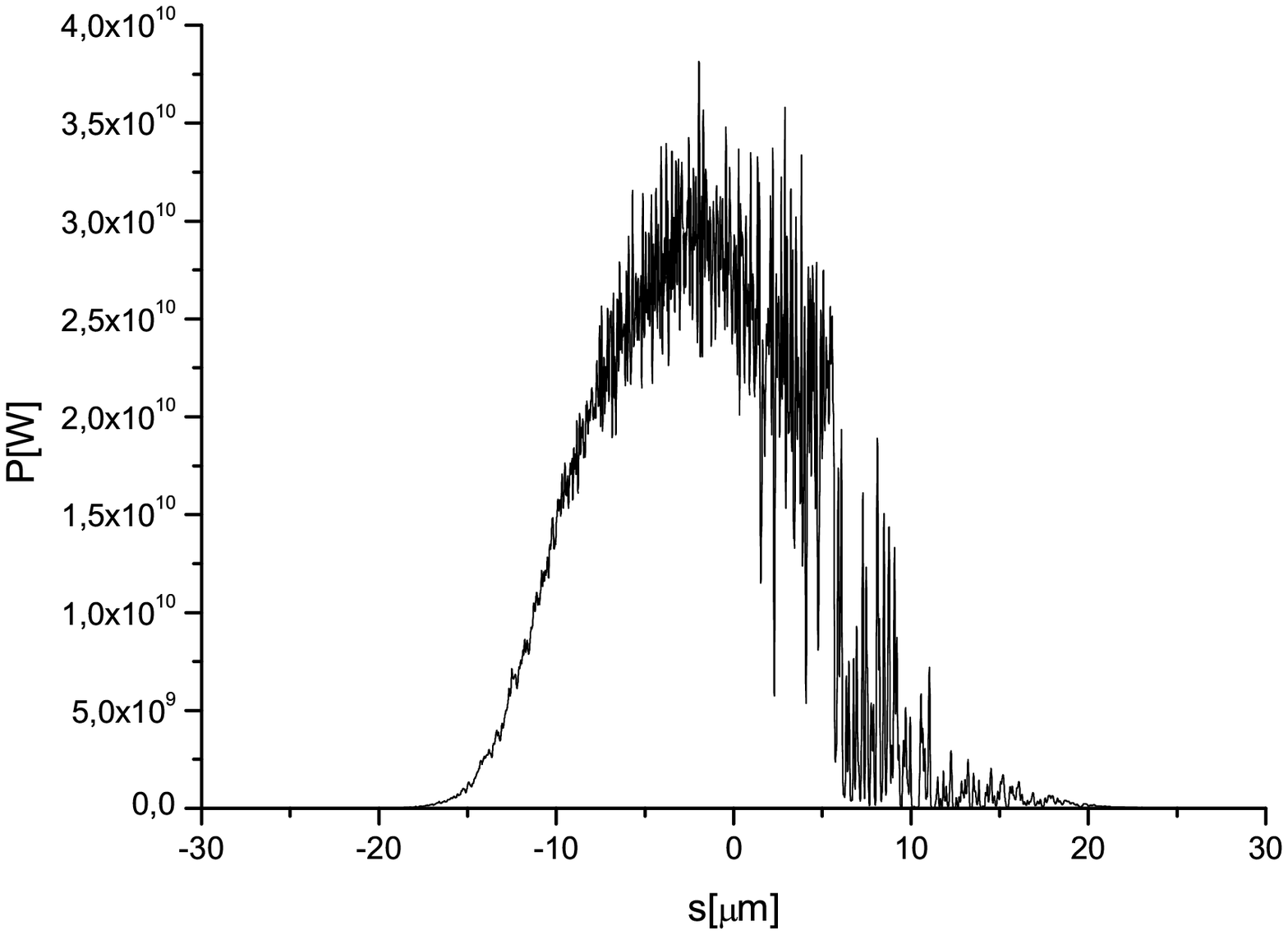}
\caption{Long pulse mode of operation, combination of self-seeding
and fresh bunch techniques. Output power of the device.}
\label{LSfinout}
\end{figure}

The final output is shown in Fig. \ref{LPfinout} and Fig.
\ref{LSfinout}, which illustrate both spectrum and power
distribution after the third undulator stage. The relative
spectral width is about $10^{-5}$. 


\section{Conclusions}

In this paper we propose a novel scheme to produce highly
monochromatic X-rays from a baseline SASE XFEL, down to a relative
bandwidth of $10^{-5}$. The key components of such scheme include
only a single crystal and a short magnetic chicane with very small
offset. A distinguishing feature of our method is that it uses a
single crystal in the transmission direction, instead of a
fixed-exit four-crystal monochromator. As a result, the X-ray
optics is extremely simple to align: it involves no
beam-recombining and no scanning of the delay. The alignment
tolerance of the crystal angle is expected to be in the range of a
fraction of mrad for fitting the Bragg reflectivity line to the
SASE XFEL radiation bandwidth.

We first illustrated the method, stressing for the first time to
our knowledge the link between the dynamical theory of diffraction
and Kramers-Kronig relations when dealing with the calculation of
the phase shift for the forward-diffracted beam in Bragg geometry.
Subsequently, we presented a feasibility study of the proposed
technique for both short and long pulse mode of operation.

A great advantage of our method is that it includes no path delay
of X-rays in the monochromator. This fact eliminates the need for
a long electron beam bypass, or for the creation of two precisely
separated, identical electron bunches, as required in previously
proposed self-seeding schemes. The present scheme is therefore
inexpensive and compact. Moreover, the proposed combination of
single crystal and weak chicane allows for a straightforward
installation of the self-seeding setup in the baseline undulator
system, already during the commissioning phase of SASE2 at the
European XFEL, and can become operational by the end of 2014.

\section{Acknowledgements}

We are grateful to Massimo Altarelli, Reinhard Brinkmann and
Serguei Molodtsov for their support and their interest during the
compilation of this work, to Martin Tolkiehn and Edgar Weckert for
useful discussions.

\end{document}